\documentclass[twoside,twocolumn,9pt]{article}
\usepackage{extsizes}
\usepackage[super,sort&compress,comma]{natbib} 
\usepackage[version=3]{mhchem}
\usepackage[left=1.5cm, right=1.5cm, top=1.785cm, bottom=2.0cm]{geometry}
\usepackage{balance}
\usepackage{mathptmx}
\usepackage{sectsty}
\usepackage{graphicx} 
\usepackage{lastpage}
\usepackage{subfigure}
\usepackage[format=plain,justification=justified,singlelinecheck=false,font={stretch=1.125,small,sf},labelfont=bf,labelsep=space]{caption}
\usepackage{float}
\usepackage{fancyhdr}
\usepackage{fnpos}
\usepackage[english]{babel}
\addto{\captionsenglish}{%
  \renewcommand{\refname}{Notes and references}
}
\usepackage{array}
\usepackage{droidsans}
\usepackage{charter}
\usepackage[T1]{fontenc}
\usepackage[usenames,dvipsnames]{xcolor}
\usepackage{setspace}
\usepackage[compact]{titlesec}
\usepackage{hyperref}

\usepackage{epstopdf}

\definecolor{cream}{RGB}{222,217,201}

\begin{document}

\pagestyle{fancy}
\thispagestyle{plain}
\fancypagestyle{plain}{
\renewcommand{\headrulewidth}{0pt}
}

\makeFNbottom
\makeatletter
\renewcommand\LARGE{\@setfontsize\LARGE{15pt}{17}}
\renewcommand\Large{\@setfontsize\Large{12pt}{14}}
\renewcommand\large{\@setfontsize\large{10pt}{12}}
\renewcommand\footnotesize{\@setfontsize\footnotesize{7pt}{10}}
\makeatother

\renewcommand{\thefootnote}{\fnsymbol{footnote}}
\renewcommand\footnoterule{\vspace*{1pt}%
\color{cream}\hrule width 3.5in height 0.4pt \color{black}\vspace*{5pt}} 
\setcounter{secnumdepth}{5}

\makeatletter 
\renewcommand\@biblabel[1]{#1}            
\renewcommand\@makefntext[1]%
{\noindent\makebox[0pt][r]{\@thefnmark\,}#1}
\makeatother 
\renewcommand{\figurename}{\small{Fig.}~}
\sectionfont{\sffamily\Large}
\subsectionfont{\normalsize}
\subsubsectionfont{\bf}
\setstretch{1.125} 
\setlength{\skip\footins}{0.8cm}
\setlength{\footnotesep}{0.25cm}
\setlength{\jot}{10pt}
\titlespacing*{\section}{0pt}{4pt}{4pt}
\titlespacing*{\subsection}{0pt}{15pt}{1pt}

\fancyfoot{}
\fancyfoot[LO,RE]{\vspace{-7.1pt}\includegraphics[height=9pt]{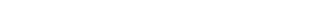}}
\fancyfoot[CO]{\vspace{-7.1pt}\hspace{13.2cm}\includegraphics{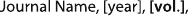}}
\fancyfoot[CE]{\vspace{-7.2pt}\hspace{-14.2cm}\includegraphics{head_foot/RF}}
\fancyfoot[RO]{\footnotesize{\sffamily{1--\pageref{LastPage} ~\textbar  \hspace{2pt}\thepage}}}
\fancyfoot[LE]{\footnotesize{\sffamily{\thepage~\textbar\hspace{3.45cm} 1--\pageref{LastPage}}}}
\fancyhead{}
\renewcommand{\headrulewidth}{0pt} 
\renewcommand{\footrulewidth}{0pt}
\setlength{\arrayrulewidth}{1pt}
\setlength{\columnsep}{6.5mm}
\setlength\bibsep{1pt}

\makeatletter 
\newlength{\figrulesep} 
\setlength{\figrulesep}{0.5\textfloatsep} 

\newcommand{\topfigrule}{\vspace*{-1pt}%
\noindent{\color{cream}\rule[-\figrulesep]{\columnwidth}{1.5pt}} }

\newcommand{\botfigrule}{\vspace*{-2pt}%
\noindent{\color{cream}\rule[\figrulesep]{\columnwidth}{1.5pt}} }

\newcommand{\dblfigrule}{\vspace*{-1pt}%
\noindent{\color{cream}\rule[-\figrulesep]{\textwidth}{1.5pt}} }

\makeatother

\twocolumn[
  \begin{@twocolumnfalse}
{\includegraphics[height=30pt]{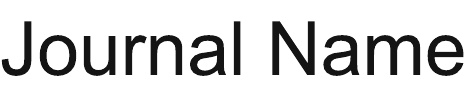}\hfill\raisebox{0pt}[0pt][0pt]{\includegraphics[height=55pt]{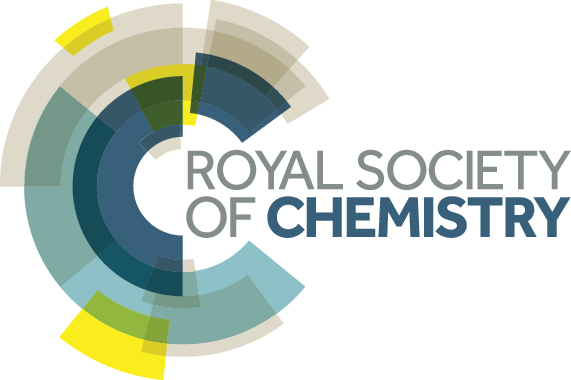}}\\[1ex]
\includegraphics[width=18.5cm]{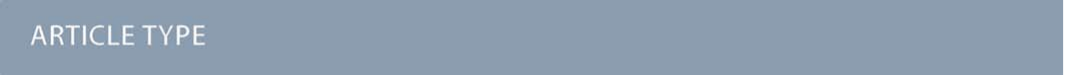}}\par
\vspace{1em}
\sffamily
\begin{tabular}{m{4.5cm} p{13.5cm} }

\includegraphics{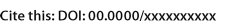} & \noindent\LARGE{\textbf{Tuning the  Electronic and Magnetic Properties of Double Transition Metal MCrCT$_2$ (M = Ti, Mo) Janus MXenes for Enhanced Spintronics and Nanoelectronics}} \\
\vspace{0.3cm} & \vspace{0.3cm} \\

 & \noindent\large{Swetarekha Ram,\textit{$^{a}$} Namitha Anna Koshi,\textit{$^{a}$}
Seung-Cheol Lee $^{\ast}$\textit{$^{a,b}$} and Satadeep Bhattacharjee $^{\ast}$\textit{$^{a}$}} \\

\includegraphics{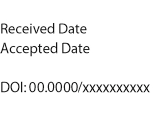} & \noindent\normalsize{Janus MXenes, a new category of two-dimensional (2D) materials, shows promising potential for advances in optoelectronics, spintronics and nanoelectronics. Our theoretical investigations not only provide interesting insights but also highlight the promise of Janus MCrCT$_2$ (M = Ti, Mo; T = O, F, OH) MXenes for future spintronic applications and highlight the need for their synthesis. Electronic structure analysis shows different metallic and semi-metallic properties: MoCrCF$_2$ exhibits metallic properties, TiCrC(OH)$_2$ and MoCrCO$_2$ exhibit near semi-metallicity with spin polarization values of 61\% and 86\%, respectively, while TiCrCO$_2$ and TiCrCF$_2$ are completely half-metallic with 100\% spin polarization at the Fermi level. All studied Janus MXenes exhibit intrinsic ferromagnetism, which is mainly attributed to the chromium (Cr) atoms, as shown by the spin density difference plots. Among them, the TiCrCO$_2$ monolayer stands out with the highest exchange constant and ferromagnetic transition temperature (T$_c$). Notably, the O-terminated Janus MXenes exhibit weak perpendicular magnetic anisotropy, in contrast to the in-plane anisotropy observed for F and OH-terminated MXenes, making them particularly interesting for future spintronic applications which we further demonstrate with micromagnetic simulation which reveal distinct current-induced switching behaviors in these Janus MXenes with different surface terminations.}\\

\end{tabular}

 \end{@twocolumnfalse} \vspace{0.6cm}

  ]

\renewcommand*\rmdefault{bch}\normalfont\upshape
\rmfamily
\section*{}
\vspace{-1cm}

\footnotetext{\textit{$^{a}$~Indo-Korea Science and Technology Center (IKST), Jakkur, Bengaluru 560065, India}}
\footnotetext{\textit{$^{b}$~Electronic Materials Research Center, KIST, Seoul 136-791, South Korea}}
\footnotetext{$^\ast$~E-mail:leesc@kist.re.kr, s.bhattacharjee@ikst.res.in}


\footnotetext{\dag~Electronic Supplementary Information (ESI) available: Phonon dispersion of MoCrC(OH)$_2$; AIMD simulation for Janus MXene; FM, AFM and the various configurations used for J calculation; Atom resolved band structure for Janus MXene. See DOI: 00.0000/00000000.}



\section{Introduction}
Recent advancements in the study of two-dimensional (2D) materials can be attributed to the discovery of graphene. Due to their reduced dimensions and sizes, 2D crystals have demonstrated a variety of fascinating features and are now regarded as the foundation of nanoscale electronics and spintronics.\cite{xu2013graphene,Si2013first,novoselov2012roadmap,wang2012electronics} 
It is difficult to meet the practical requirements of microelectronic devices with the few known 2D intrinsically magnetic materials because of their low Curie temperature (50 K).\cite{zhang2013intrinsic,gong2017discovery} 
As a result, developing novel 2D magnetic materials has become a popular area of study.
The majority of 2D materials are inherently nonmagnetic in their unaltered states, which restricts their use in spintronic devices. Low dimensional systems are well suited for a variety of applications because of the tunability of their electronic and magnetic properties. Numerous methods are reported to append magnetism in 2D materials, which includes doping, point defects, and atom adsorption\cite{babar2016transition,zheng2014tuning}; applying an external electric field and tensile strain;\cite{zhou2012tensile} depositing magnetic foreign atoms on the material; \cite{sevinccli2008electronic} and incorporating specific defects or edges to the material.\cite{avsar2019defect,zhang2013intrinsic} In addition to being excellent candidates for high-performance spin electronic devices like ultrafast non-volatile electric/magnetic memory devices, magnetic materials with coupled electric and magnetic orders also offer a fantastic opportunity to realise the magnetic control of electric polarization or electrical control of magnetization.

In recent times, a new class of 2D materials\cite{naguib2011two} known as "MXene" has drawn a lot of attention because of its exceptional properties, which include high conductivity and hydrophilic behaviour, application in electrode materials, sensors, catalysis and electrochemical energy storage, and electrocatalysis.\cite{naguib2012two,ling2014flexible,lukatskaya2013cation,eames2014ion,zhang2018single} The development of spintronic devices require a number of critical features, like high spin polarization, high Curie temperature, (T$_c$) and substantial magnetocrystalline anisotropy energy. Because the majority of them are nonmagnetic, there aren't many reports on MXene's magnetic properties. Based on Monte Carlo simulations of the traditional 2D Heisenberg model, Sun et al.\cite{sun2021cr2nx2} evaluated the Curie temperature of Cr$_2$NT$_2$, and their findings suggested that the functionalized Cr$_2$NT$_2$ would be of major significance for creating efficient spintronic devices for room-temperature applications.\cite{he2016high}  The intrinsic half-metal Mn$_2$CF$_2$ MXene is predicted to have a large half-metallic gap (0.9 eV) and a high Curie temperature (520 K).\cite{he2016new}  It has been hypothesised that Mn$_2$CT$_2$ (T = F, Cl, or OH) and Cr$_2$NT$_2$ (T = O, F, OH)\cite{sun2021cr2nx2} are half-metallic compounds.  Khazaei et al.\cite{khazaei2013novel} were the first to describe the magnetism of -F and -OH functionalized Cr-based MXenes (Cr$_2$CF$_2$ and Cr$_2$C(OH)$_2$). Si et al.\cite{si2015half} expected the bare Cr$_2$C MXene to be half-metallic and they computed that the -F, -OH, chlorine (Cl), and hydrogen (-H) functionalized Cr$_2$C were antiferromagnetic semiconductors.
 Depending on the choice of second metal atom (M$^{'}$)  and termination atom (T),  double transition metal Cr$_2M^{'}$C$_2$T$_2$  can be found as FM, AFM, or NM  and either a metal or semiconductor. \cite{yang2016tunable}
For Ti$_2$C monolayer, a phase transition from roughly half-metallic to a half-metallic spin gapless semiconductor has been described.
\cite{gao2016monolayer}

A singular opportunity exists to introduce novel magnetic phenomena into spintronic devices for information and communication technology by incorporating Janus features into the magnetic monolayer. This hybrid system will retain the intrinsic properties of a single 2D material, even though interfacial contact may lead to the formation of new physical properties.
Two-dimensional (2D) Janus structures that have been experimentally realised include MoSSe\cite{Panigrahi2021two} and Janus graphene\cite{Zhang2013Janus}. These structures have drawn attention from all around the world for their new characteristics and intriguing behaviours. The electronic properties of two-dimensional materials are usually determined by the breaking of structural symmetry, as is the case with 2D Janus MXenes, which have different endpoints on each of their two surfaces. Symmetry breaking of graphene in the plane has been the subject of numerous studies (see e.g. Zhang et al.\cite{zhang2009direct} and references therein).
Also, it has been demonstrated that an additional degree of freedom allowing spin manipulation can be obtained by breaking the out-of-plane mirror symmetry, either with external electric fields,\cite{yuan2013zeeman,wu2013electrical} or with an asymmetric out-of-plane structural configuration,\cite{cheng2013spin} similar to the one found in 2D Janus MXenes. Therefore, introducing such asymmetry can result in a wide range of physical phenomena that are absent in ordinary MXenes. In their individual bare cases, the Cr-based 2D Janus MM$^{'}$C MXenes (M $\neq$ M$^{'}$ = Cr, Ti, Sc, and V) do not exhibit any peculiar electronic or magnetic properties. 
\cite{akgencc2021tuning} It becomes vital to look for substitute transition metal (TM) atoms in order to design Janus MXenes for next-generation devices. Two-dimensional (2D) ferromagnetic materials with high spin-polarization are in high demand for spintronic devices. However, there are still questions regarding the electronic structure and magnetic characteristics of 2D Janus magnetic materials with strong spin polarisation. Having 100 $\%$ spin polarization, half-metallic materials are conductive in one spin channel but insulating in the other. Any spin-based device would greatly benefit from this property. We are driven to analyse and investigate since, in contrast to semiconductor Janus structures, half metallic (HM) Janus structures that have not yet received sufficient attention. The ideal half-metal used in spintronics is projected to have a higher T$_c$. Furthermore, the enormous magnetic anisotropy energy (MAE) is crucial for magnetoelectronics.\cite{guan2020strain} In the current study, we seek to introduce the novel Janus MXenes, with half-metallic character that may hold promise as an emerging spintronic material.

 Despite the absence of Janus 2D materials in nature, Janus monolayers of transition metal dichalcogenides have also been successfully synthesised. 
 In light of the successful experimental synthesis of 2D Janus MoSSe,\cite{Panigrahi2021two} we carry out the investigation of electronic structure and magnetic properties of Janus MXene M(M=Ti,Mo)CrC functionalized with T(T = -F, -OH, and -O) atoms (MCrCT$_2$). The electronic, magnetic, and tunable properties of MXenes are mostly influenced by the surface functionalization. These motivations have led us to explore the structural, electronic, magnetic  and thermodynamic characteristics of Janus MXenes. Although it is still a challenging undertaking for the experimental community, it is fair to think that Janus MXenes can also be made experimentally, similar to Janus graphene.\cite{Zhang2013Janus}  

\section{Computational details}
The Vienna ab initio simulation package (VASP),\cite{kresse1993ab} which implements the spin-polarized density functional theory (DFT), is used for all calculations. The electron-electron interaction is approximated using the Generalised Gradient Approximation (GGA) in Perdew-Burke-Ernzerhof (PBE)\cite{perdew1996generalized} form, and the electron-ion potential is approximated using the Projector-Augmented Wave (PAW) method.\cite{blochl1994projector} To prevent interaction between neighbouring layers, a vacuum separation of 15 {\AA} is added along the z-axis. The plane wave cutoff in all computations is 500 eV. The 10$^{-6}$  eV criterion is used for the energy (stress) convergence. We employed Grimme's empirical correction approach (DFT-D3), which has been shown to be reliable for characterising the long-range van der Waals (vdW) interactions.\cite{grimme2006semiempirical} The dipole correction is taken into account in the calculations since Janus MXenes have an asymmetric structure. Their dynamic stability are examined using frozen phonon method implemented in VASP in conjunction with Phonopy code. \cite{togo2015first}
 Here, we used supercell of size 5 $\times$ 5 $\times$ 1 to calculate the force constant matrix, whose Brillouin zone is sampled with 3 $\times$ 3 $\times$ 1 k-mesh. To guarantee a satisfactory convergence, we have additionally increased the electronic degrees of freedom to 10$^{-8}$ eV. 
 To confirm structural thermodynamical stability, ab initio molecular dynamics (AIMD) simulation is additionally carried out. At temperatures of 300 K, a Nos\'{e} Hoover thermostat is used with the constant moles-volume-temperature (NVT) ensemble. The entire time is 10 ps, and the time step is 1 fs. A convergence of 10$^{-7}$ eV for the total energy is used to compute the magnetic anisotropy energy (MAE). In the MAE computation, the spin-orbit coupling (SOC) is also taken into account, and the corresponding $k$grid 12 $\times$ 12 $\times$ 1 is employed  and with no symmetry constraints. 





\begin{figure*}[h]
  \centering
     \includegraphics[scale=0.6]{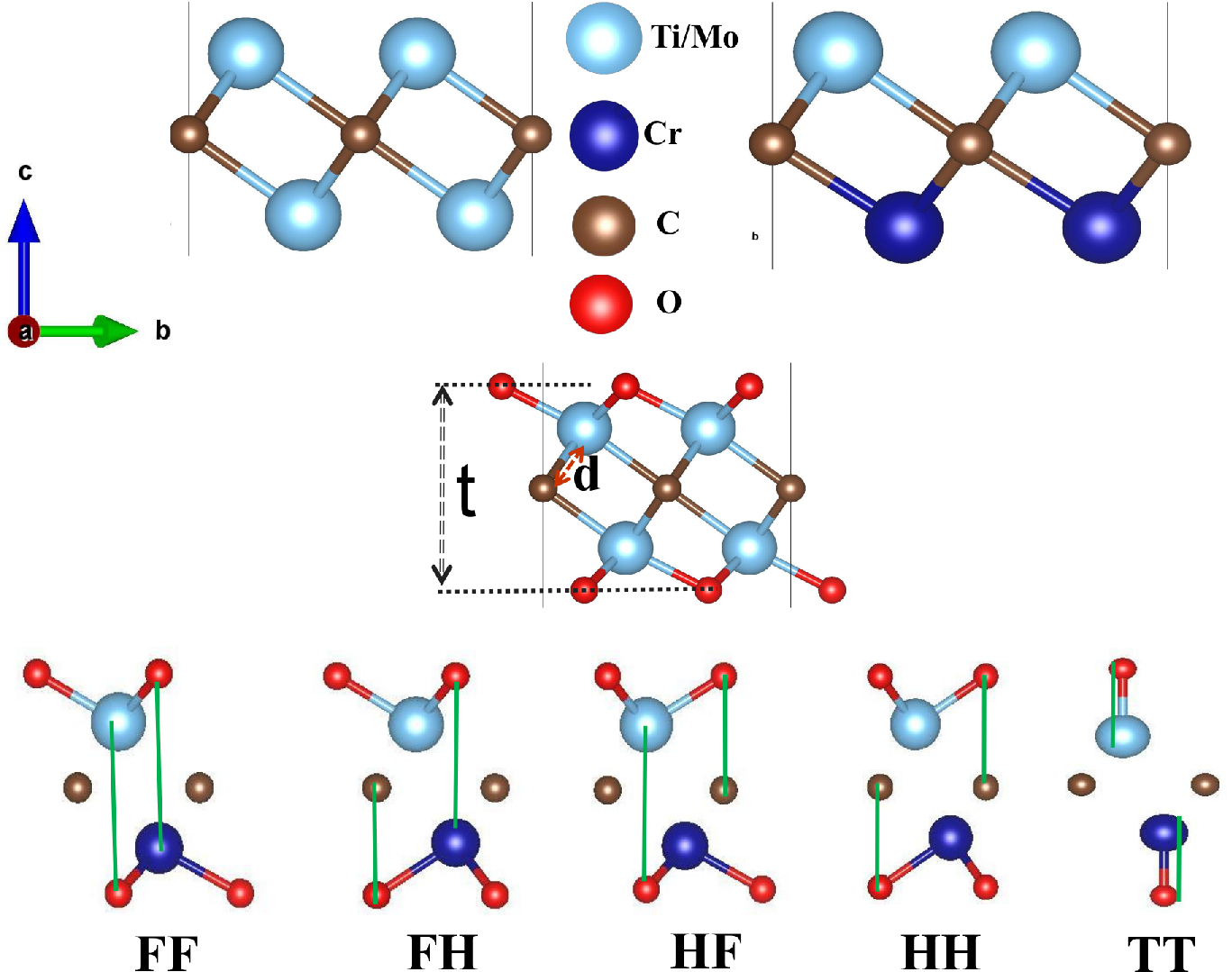}        
 \caption{ Side views of the M$_2$C, MCrC and M$_2$CO$_2$  MXene with M = Ti/Mo.  Side views of the five possible models for the termination group of  MCrCT$_2$ (T = O, F, OH) MXenes.  Here, we have demonstrated the -O termination group. The scenario is the same for other termination groups as well.}
 \label{Fig:str1}
\end{figure*}

\begin{table*}
\caption{ Lattice parameter a (\AA), thickness, t(\AA) and bond length, $d$, as well as the formation energy, H$_f$ of Janus MXene}
 \begin{tabular*}{\textwidth}{@{\extracolsep{\fill}}llllllll}
\hline
       & a    & t      & d(M-C) & d(Cr-C) & d(M-T) & d(Cr-C)  & H$_f$  \\ 
    & (\AA)& (\AA)  & (\AA)      & (\AA)   &  (\AA)     & (\AA)    & (eV)   \\ 
    \hline
TiCrCO$_2$    & 2.98 & 4.36    & 1.3        & 1.2     & 0.95       & 0.9  & -8.77  \\ 
TiCrCF$_2$    & 3.04 & 4.6      & 1.17       & 0.97    & 1.24       & 1.22 & -9.55  \\ 
TiCrC(OH)$_2$ & 3.07 & 6.56  & 1.16       & 0.97    & 1.26       & 1.22 & -12.65  \\ 
MoCrCO$_2$    & 2.91 & 4.73  & 1.3        & 1.27    & 1.22       & 0.93 & -7.77  \\ 
MoCrCF$_2$    & 3.00 & 4.98  & 1.21       & 1.03    & 1.5        & 1.23 & -8.06  \\ \hline
\end{tabular*}
\label{Table:str}
\end{table*}

\section{Results and discussions}
\subsection{Structural properties and stability}
 As seen in Figure-\ref{Fig:str1}, M$_{2}$C has a sandwich-like structure with one carbon layer sandwiched between two M (M = Ti/Mo) layers. Figure-\ref{Fig:str1} illustrates how Janus MXenes are created by removing one layer of TM from M$_2$C MXene.  According to previous results, functional groups are unavoidably formed on the bare MXenes.\cite{khazaei2013novel}. A functional group termination in MXene is possible at the fcc, hcp, or atop sites. As illustrated in Figure-\ref{Fig:str1}, the functionalized MXenes can be stacked in five different configurations: fcc-fcc (FF), fcc-hcp (FH), hcp-fcc (HF), hcp-hcp (HH), and atop-atop (TT). We completely optimized the structural models for each MCrCT$_2$ system in order to identify the most stable configuration with termination group. The results of our examination of the optimum configuration with the lowest energy are summarised in Table-\ref{Table:str}. After spin-polarized geometry optimization, the ground state of each Janus MXene has hexagonal symmetry with P3m1 space group. Due to the broken central symmetry, the point group of MCrCO$_2$ monolayer is C$_{3v}$ as opposed to D$_{3d}$ of M$_2$CO$_2$ monolayer. 

\begin{figure*}
  \centering
     \subfigure[]{\includegraphics[scale=0.3]{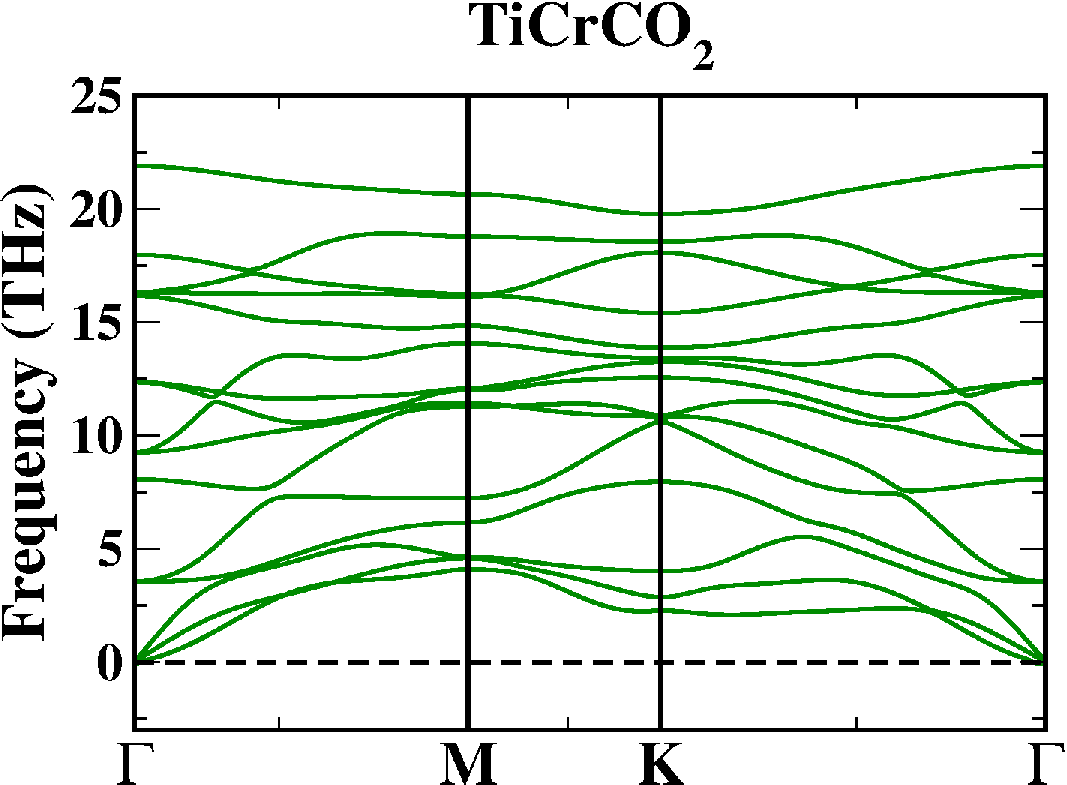}} 
     \subfigure[]{\includegraphics[scale=0.3]{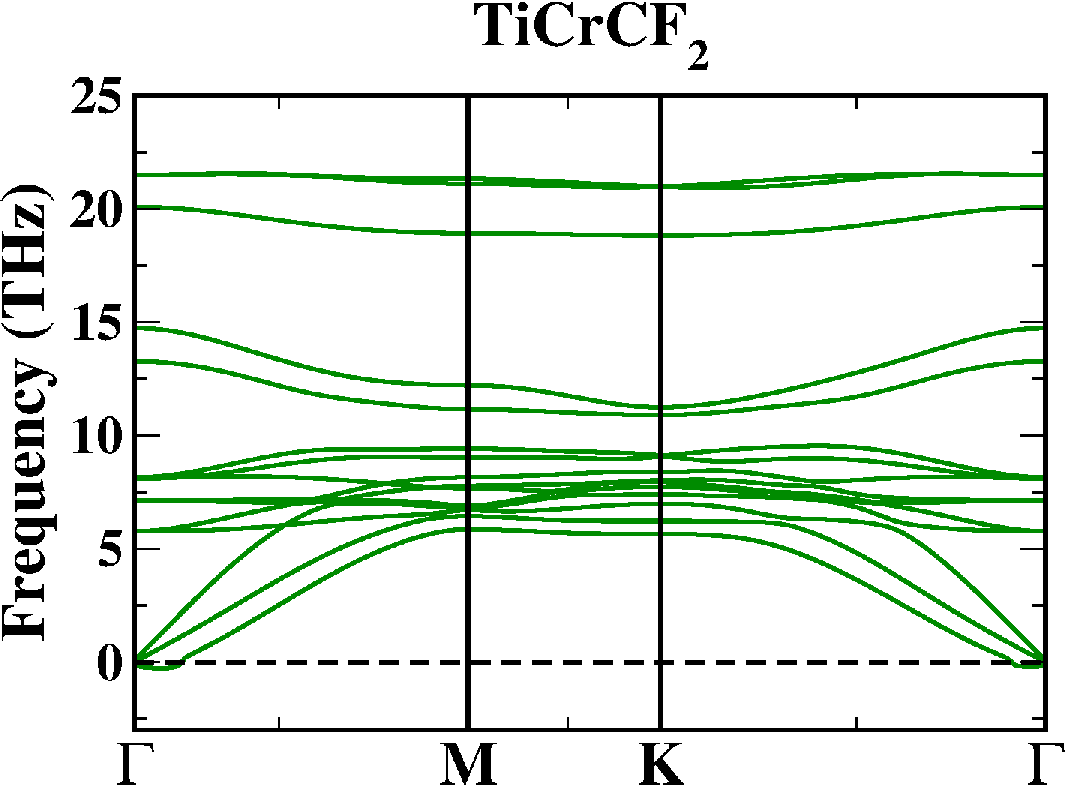}}  
     \subfigure[]{\includegraphics[scale=0.3]{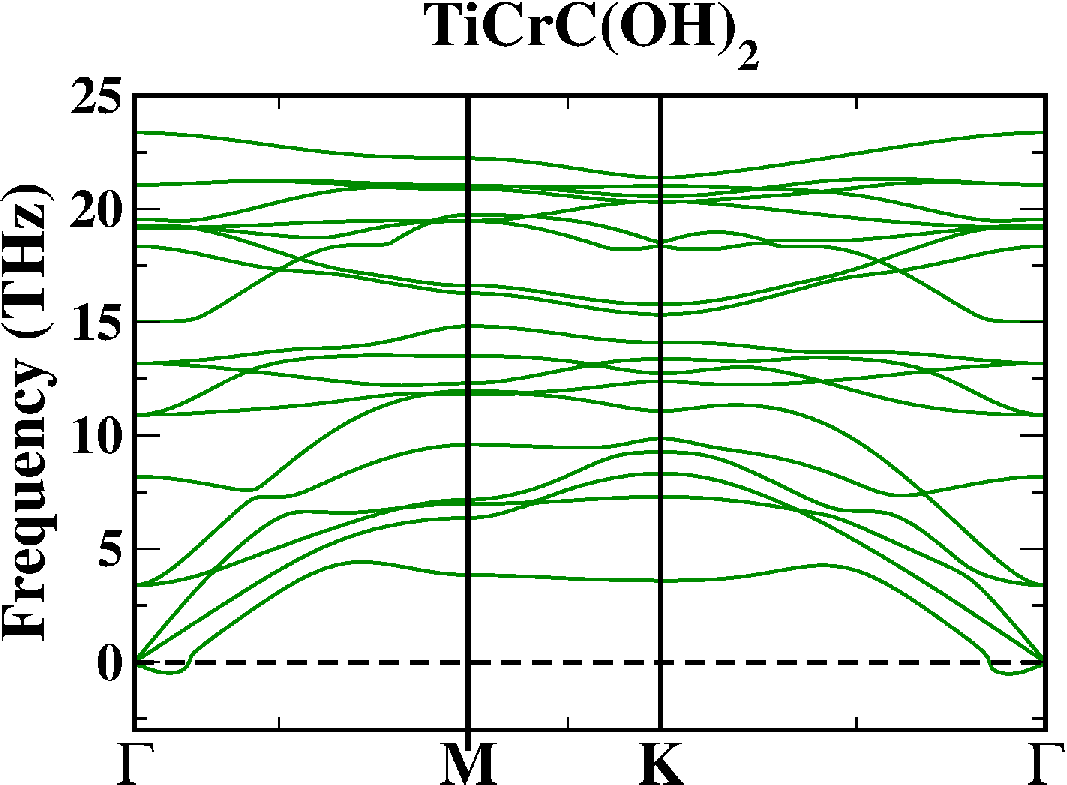}}  
     \subfigure[]{\includegraphics[scale=0.3]{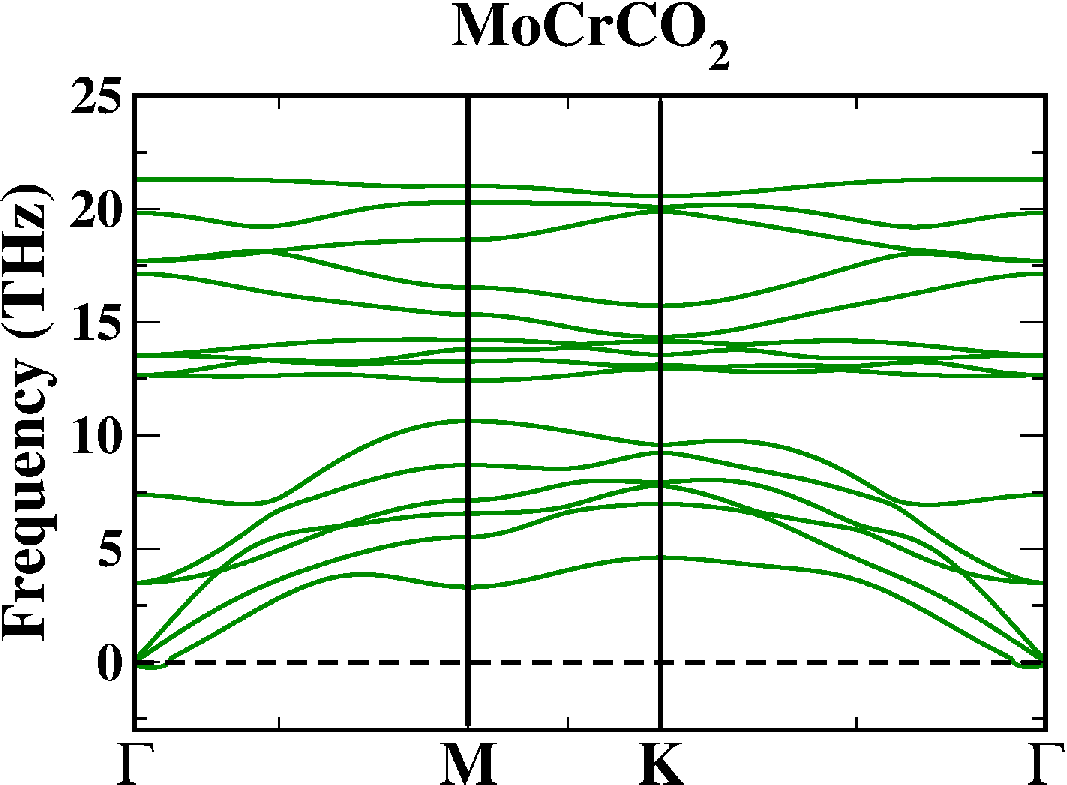}}  
     \subfigure[]{\includegraphics[scale=0.3]{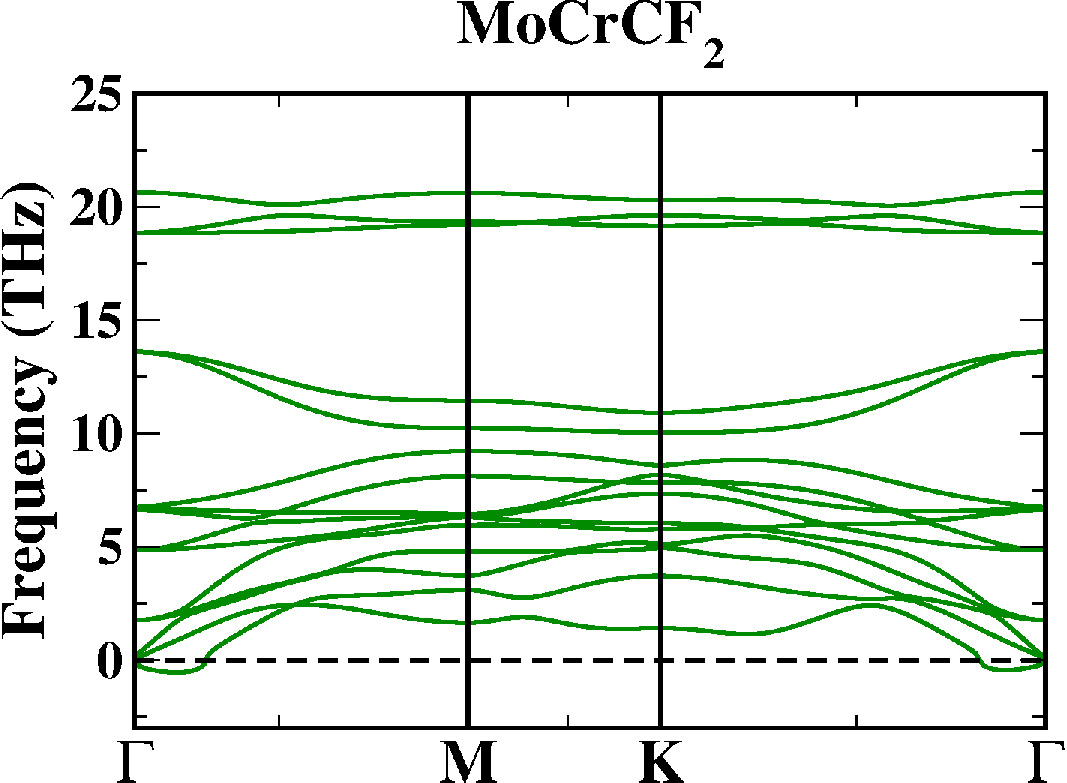}}

 \caption{Phonon band structure of (a) TiCrCO$_2$, (b) TiCrCF$_2$, (c) TiCrC(OH)$_2$, (d) MoCrCO$_2$, (e) MoCrCF$_2$, respectively. The figures provide evidence that MCrCT$_2$ Janus structures are dynamically stable. }
 \label{Fig:phonon}
\end{figure*}

The relaxed lattice constant (a) and thickness, t, of surface terminated MCrCT$_2$ Janus MXenes are listed in Table-\ref{Table:str}. Formation energies are calculated to get the most stable configuration based on
following relation 
\begin{equation}
 \Delta H_f = E_{tot} (MCrCT_2) + E_{tot}(MCrC) + E_{tot}(T_2)    
\end{equation}

where $E_{tot}$ (MCrCT$_2$), $E_{tot}$(MCrC) and $E_{tot}$ (T$_2$) are the total energy of fully surface terminated MCrCT$_2$, total energy of MCrC, and total energy of T$_2$ (F$_2$, O$_2$ or (OH)$_2$) in gaseous state, respectively. Table-\ref{Table:str}  contains the predicted formation energies of completely surface-terminated Janus MXene. The dynamical stability of MCrCT$_2$ MXene (phonon band dispersions shown in Figure-\ref{Fig:phonon}) is evaluated using the phonon spectra computations prior to determining the electronic structure and magnetic properties of monolayer MXenes.  Many 2D materials and their Janus counterparts, such as graphene\cite{bonini2007phonon} and Cr$_2$I$_3$X$_3$,\cite{zhang2020spin} have minuscule imaginary frequencies at the $\Gamma$ point that are analogous to the cases discussed here. A larger supercell could remedy this,\cite{csahin2009monolayer} providing evidence of dynamic stability. We only see imaginary frequency (Figure-S1) in MoCrC(OH)$_2$, which is not considered for further investigation. The evolution of total energy over time is computed using AIMD to support the thermal stability. The frameworks of MCrCT$_2$  are well conserved during the simulation period, and the total energy shows low fluctuation with increasing time (Figure-S2), demonstrating the thermal stability of MCrCT$_2$.

\subsection{Electronic and Magnetic properties}
\subsubsection{Density of States, Spin Charge Density and Electro-static Potential}
\begin{figure*}
  \centering
  \subfigure[]{\includegraphics[scale=0.3]{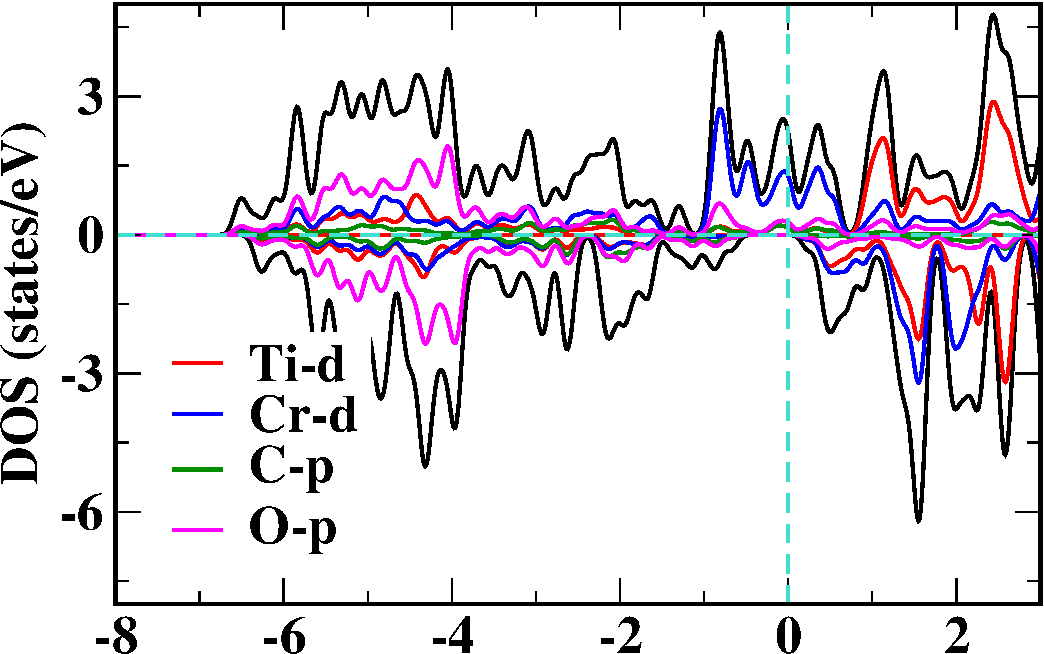}}
     \subfigure[]{\includegraphics[scale=0.3]{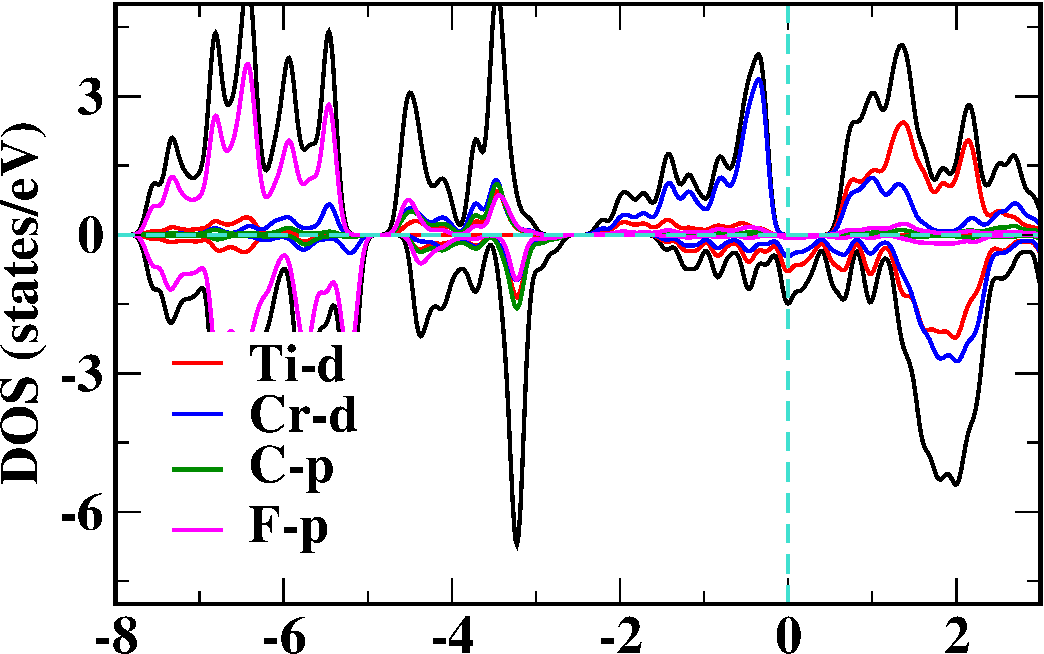}}
     \subfigure[]{\includegraphics[scale=0.3]{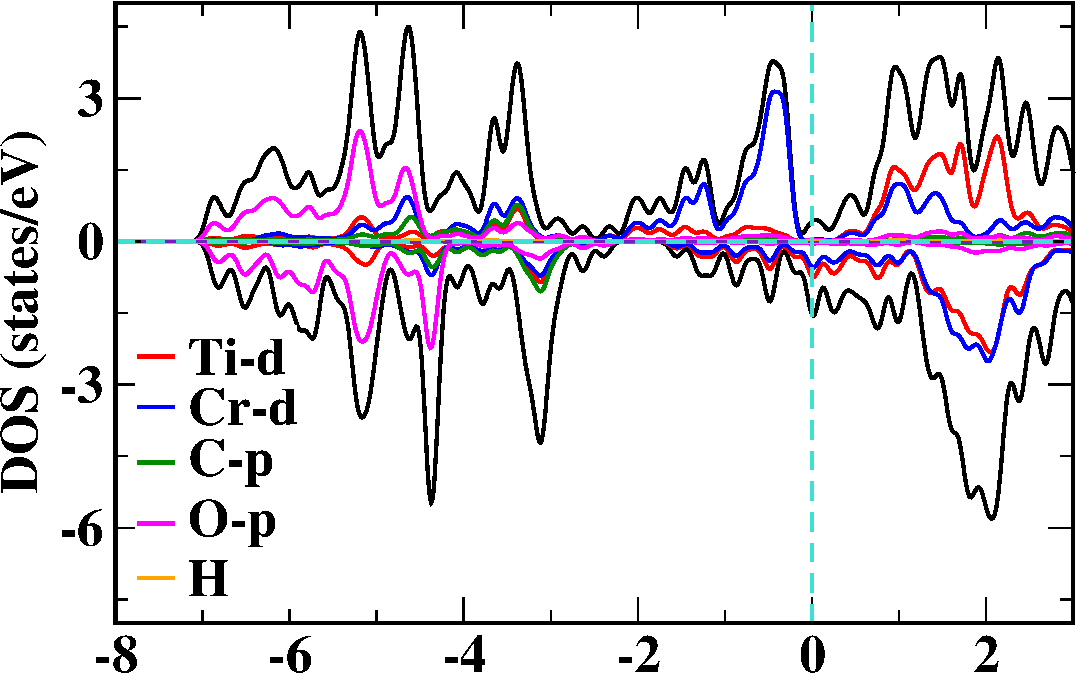}}
     \subfigure[]{\includegraphics[scale=0.3]{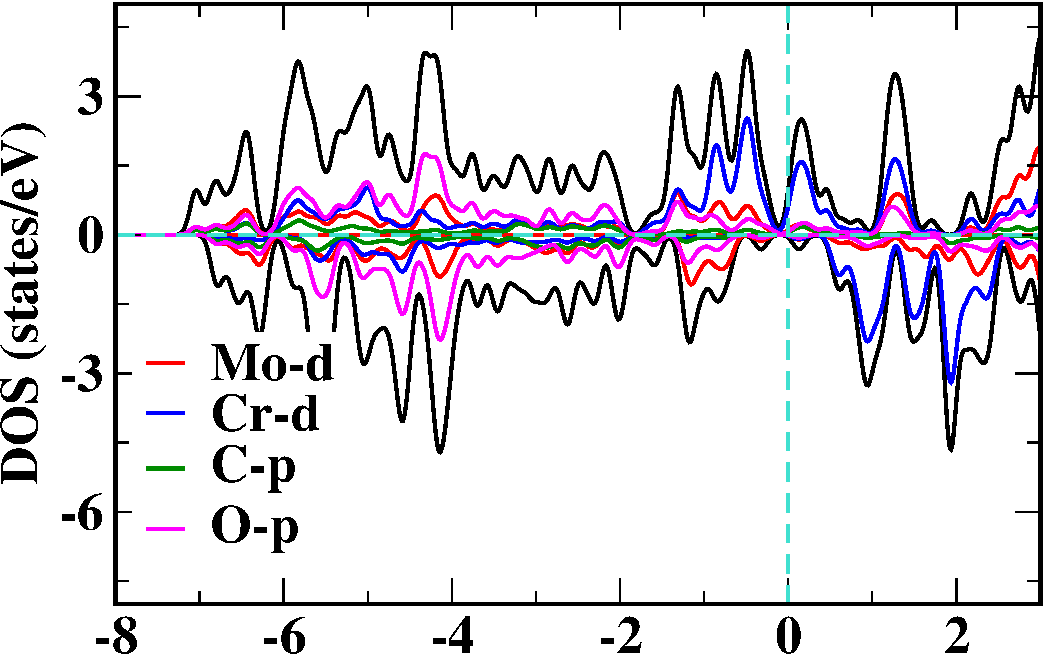}}
     \subfigure[]{\includegraphics[scale=0.3]{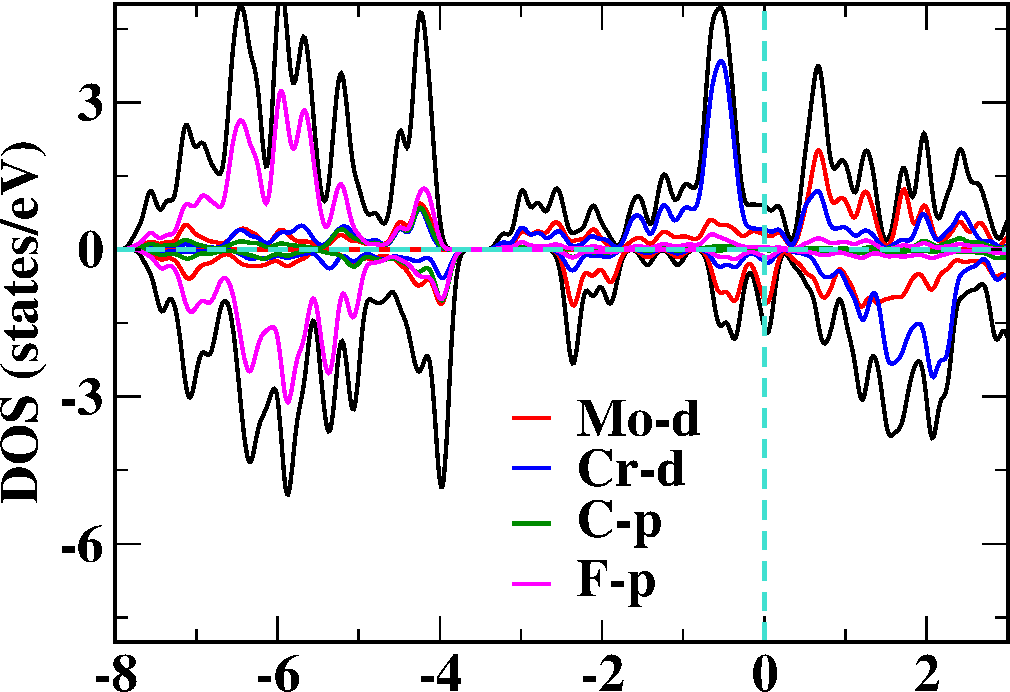}}\\
 \caption{Spin-resolved partial density of states  of (a) TiCrCO$_2$, (b) TiCrCF$_2$, (c) TiCrC(OH)$_2$, (d) MoCrCO$_2$, (e) MoCrCF$_2$, respectively. 
 }
 \label{Fig:band-dos}
\end{figure*}

\begin{figure*}
  \centering
         {\includegraphics[scale=0.5]{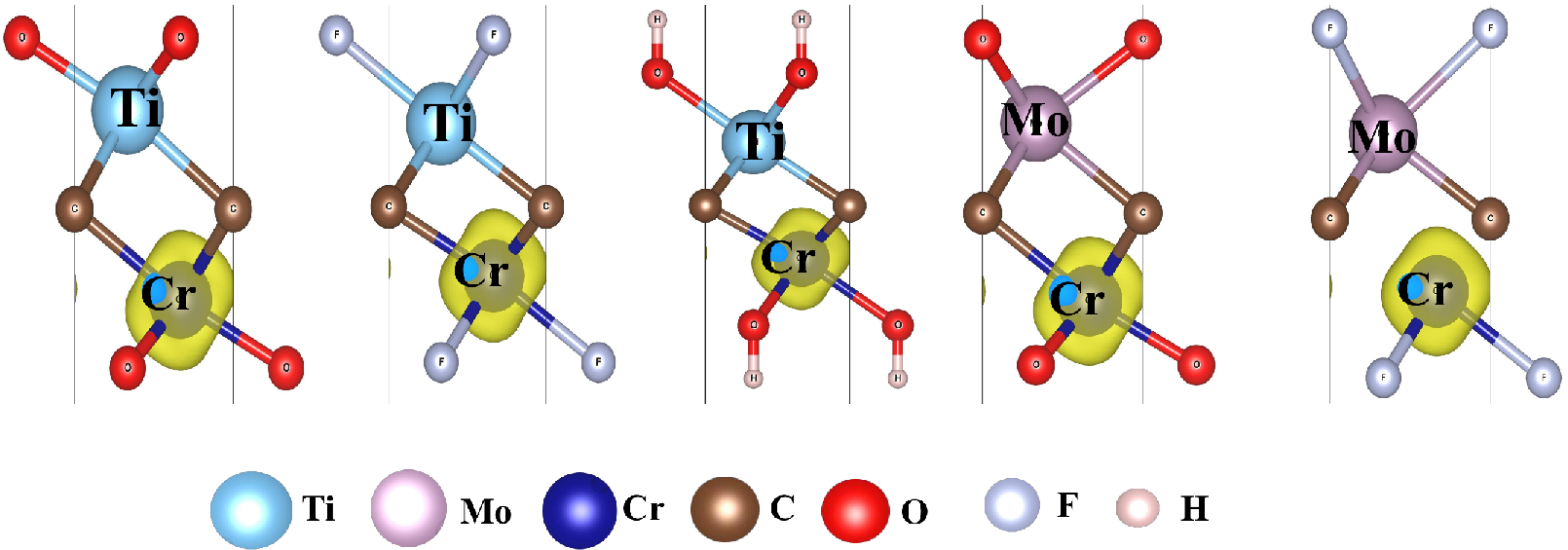}}
      \caption{ Spin charge density difference of (a) TiCrCO$_2$, TiCrCF$_2$, TiCrC(OH)$_2$, MoCrCO$_2$, MoCrCF$_2$. }
 \label{Fig:magnotation}
\end{figure*}

\begin{figure*}
  \centering
     \subfigure[]{\includegraphics[scale=0.55]{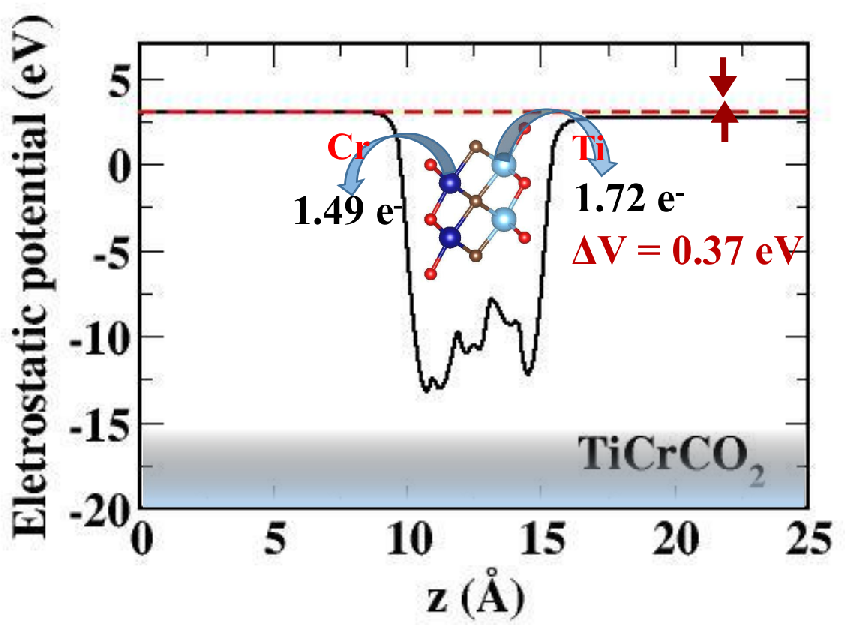}}
     \subfigure[]{\includegraphics[scale=0.5]{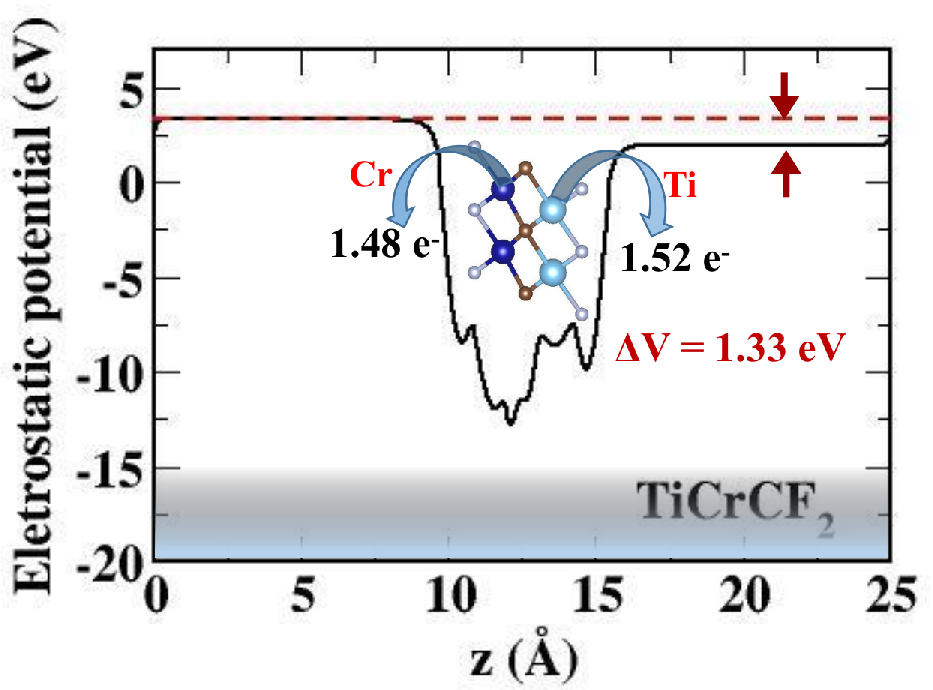}}
     \subfigure[]{\includegraphics[scale=0.58]{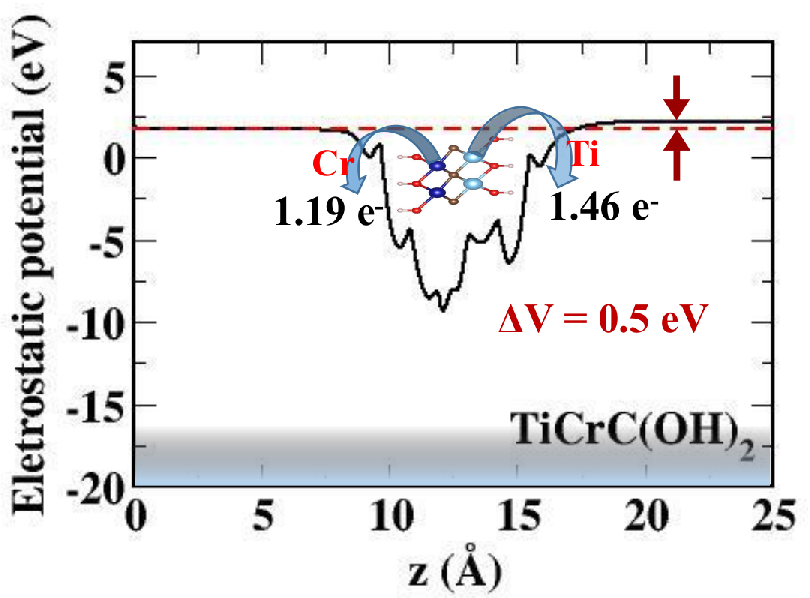}}\\
     \subfigure[]{\includegraphics[scale=0.58]{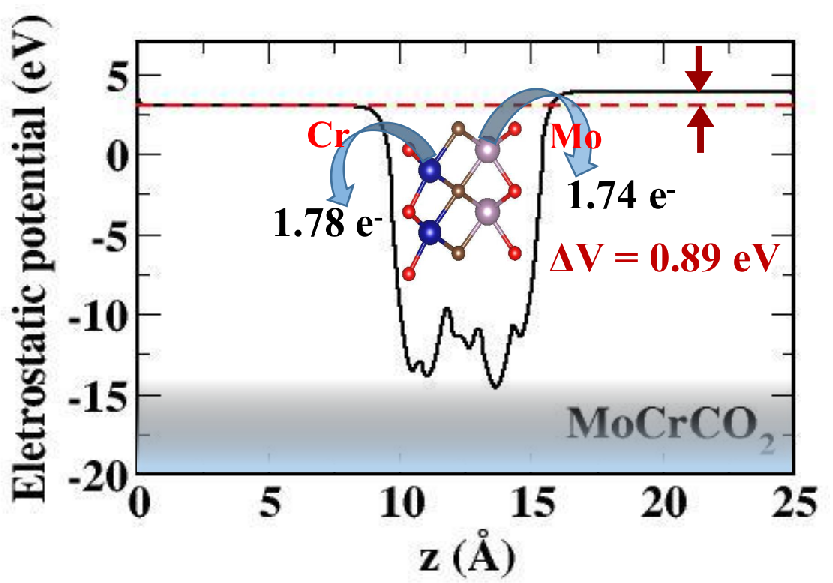}}
     \subfigure[]{\includegraphics[scale=0.57]{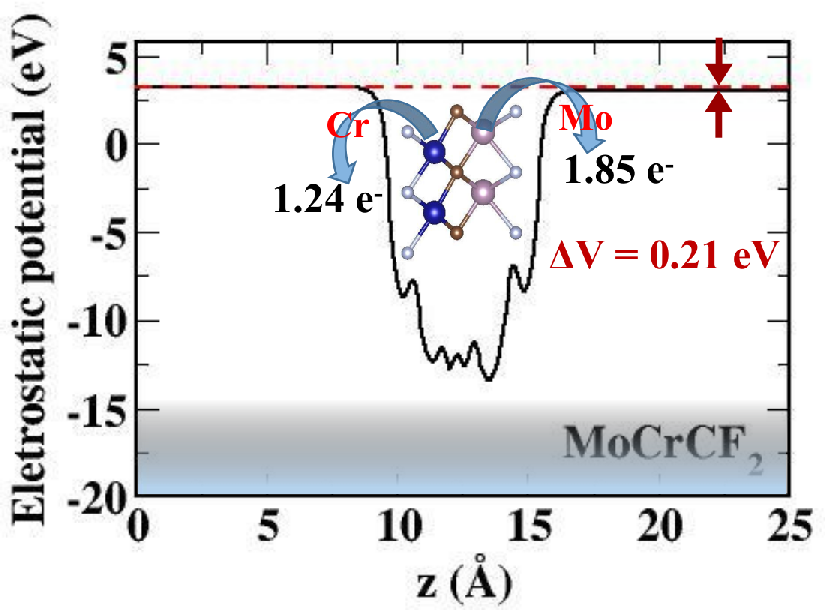}}       
 \caption{Electrostatic potential of (a) TiCrCO$_2$, (b) TiCrCF$_2$, (c) TiCrC(OH)$_2$, (d) MoCrCO$_2$, (e) MoCrCF$_2$. $\Delta V$ is the potential difference represented by red arrow mark. Bader charge are given for Cr and M atoms }
 \label{Fig:pot}
\end{figure*}

The prospective applications of 2D magnetic materials depend significantly on their electronic and magnetic characteristics.
Designing a half-metallic ferromagnet with 100\% spin-polarization at the Fermi level is conceivable when taking into account the asymmetric transition metal layers in MCrCO$_2$ (M = Ti/Mo) MXene. Spin polarization (P) is defined as P = $\left |\frac{N_\uparrow(E_F)-N_\downarrow(E_F)}{N_\uparrow(E_F)+N_\downarrow(E_F)}  \right |$, where the spin-up density of states (DOS) at the Fermi level and the spin-down density of states, respectively, are denoted by $N_\uparrow(E_F)$ and $N_\downarrow(E_F)$. The calculated values are tabulated in Table-\ref{tab:mag}. In accordance with the DOS plots in Figure-\ref{Fig:band-dos}, our computed results indicate that MoCrCF$_2$ is metallic, TiCrC(OH)$_2$ and MoCrCO$_2$ are semi-metallic, and TiCrC(OH)$_2$ and TiCrCF$_2$ are fully half metallic. Additionally, the total spin polarization indicates that MCrCT$_2$ monolayers may be used in spintronic devices, such as pure spin injection and spin transport. The non-zero magnetic moment in Table-\ref{tab:mag} and the asymmetric partial DOS (PDOS) in Figure-\ref{Fig:band-dos} imply that the Ti/Mo atoms appear to be spin split and that the Cr atom is the primary source of magnetism. PDOS simultaneously reveals that the exchange splitting is greater in the case of Cr and is insignificant in the case of Ti and Mo. As a result, Cr has a significant effect on the total magnetic moment. The difference in the corresponding spin density is again shown in Figure-\ref{Fig:magnotation}(a), where, the spin  density difference ($\Delta \rho$) is described by the equation: $\Delta \rho=\rho_{\uparrow} - \rho_{\downarrow} $ , and  $\rho_{\uparrow}$ and $\rho_{\downarrow}$ are the spin-up and spin-down electron densities, respectively.  In other words, Figure-\ref{Fig:magnotation}(a)) confirms that the magnetic moment in Cr atoms is mainly restricted. Figure-\ref{Fig:pot} depicts the electrostatic potential relationship of 2D Janus MXenes.  It is evident from the examination of the electrostatic potential diagram that the non-uniformity of charge results in a potential difference. The Janus structure's internal electrostatic potential change is represented by the valley. For the purpose of elucidating the physical cause of polarization, the Bader charge analysis of the Janus monolayer in Figure-\ref{Fig:pot} can quantitatively provide the accumulation of electrons on M(Ti/Mo) and Cr.

\begin{table}
\caption{The spin-polarisation,P, total magnetic moment, $\mu_{tot}$, local magnetic moment of M (M = Ti/Mo), $\mu_M$ and Cr, $\mu_{Cr}$ as well as local magnetic moment of functional groups(T1 and T2) in $\mu_B$, where T1 and T2 corresponds to the functional group bound to M and Cr, respectively. }
\begin{tabular}{lllllll}
\hline
           & P  & $\mu_{tot}$ & $\mu_M$ & $\mu_{Cr}$ &$\mu_{T1}$ &  $\mu_{T2}$    \\ 
           \hline

TiCrCO$_2$    &100 & 2 & 0.036 & 2 & 0.008 & -0.026 \\ 
TiCrCF$_2$    &100 &2  & -0.29 & 2.285 & -0.005    & 0.039  \\ 
TiCrC(OH)$_2$ &61 & 2.08 & -0.195  & 2.24 & -0.001 & 0.023 \\ 
MoCrCO$_2$    & 86 & 2.11 & 0.057 & 2.139  & 0.008 & -0.049 \\ 
MoCrCF$_2$    & 32 & 2.32 & -0.235 & 2.542 & -0.036 & 0.062 \\ \hline
\end{tabular}
\label{tab:mag}
\end{table}

\subsubsection{Band Structure, Fermi Surface and Electrical Conductivity}
In most of the previous reports, a rough assessment of electrical conductivity of MXenes is carried out by evaluating the density of states at the Fermi level. This is not sufficient as such an evaluation does not take into account the effect of Fermi velocity on conductivity. Recently, M. Bagheri et al. discussed a more effective treatment of electrical conductivity from first principles calculations \cite{bagheri2021fermiology} by including Fermi velocity. Using this method, we calculate the film conductivity of Janus MXenes investigated in this report. Since we employ unit cells for our calculations, the determination of band structure and Fermi surface is pretty straightforward and the issue of band unfolding does not arise. Fermi surfaces are obtained using IFERMI software \cite{ganose2021ifermi} with a large interpolation factor.

Within the constant relaxation time approximation, the sheet conductivity of a 2D material can be calculated with the knowledge of average Fermi velocity and Fermi surface area. The detailed methodology is presented by M. Bagheri et al. \cite{bagheri2021fermiology} in their supplemental material. The final expression for sheet conductivity of MXene is given by 
\begin{equation}
	(\frac{\sigma_{2D}}{\tau_{0}}) = \frac{e^{2}}{2\pi^{2} \hbar} \frac{\langle v_{F} \rangle}{2} l_{F}
 \end{equation}

 where $\hbar$ is the reduced Planck's constant, l$_{F}$ the Fermi surface length (l$_{F}$ = $\frac{a_{F}}{b_{z}}$ and b$_{z}$ = $\frac{2\pi}{L_{z}}$: a$_{F}$, Fermi surface area and L$_{z}$, length of unit cell in the z-direction) and $\langle v_{F}\rangle$ is the average Fermi velocity. The conductivities obtained in this way cannot be compared with the experimentally measured ones as they are for MXene films not monolayers. The film conductivity can be calculated using $\sigma$ = ($\frac{\sigma_{2D}}{\tau_{0}})\times(\frac{\tau_{0}}{d}$). Here, $\tau_{0}$ is the relaxation time and d is the interlayer spacing. For all calculations, we are assuming $\tau_{0}$ to be 1 fs as it gives the value of conductivity close to experimental reports \cite{bagheri2021fermiology}. The value of d taken from experiments are 7.7 Å \cite{wang2015pseudocapacitance} for Ti$_{2}$CT$_{z}$ and 9.16 Å \cite{yi2022selenium} for Mo$_{2}$CT$_{z}$. These values are used in our calculations on Janus MXenes (\ce{TiCrCT2} and \ce{MoCrCT2}, T = F,  O, OH)) as experimental data on their interlayer separations are not available. 

\begin{figure*}
\centering
\subfigure[$ $]{\hspace{-0.8cm}\includegraphics[trim=0mm 4mm 0mm 3mm,clip,scale=0.62]{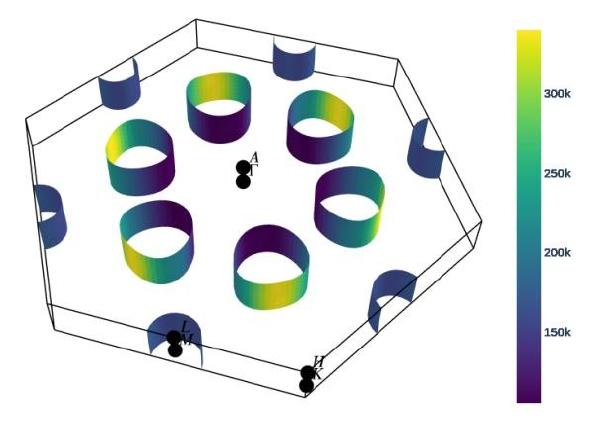}} 
\subfigure[$ $]{\hspace{0.5cm}\includegraphics[trim=0mm 4mm 0mm 5mm,clip,scale=0.61]{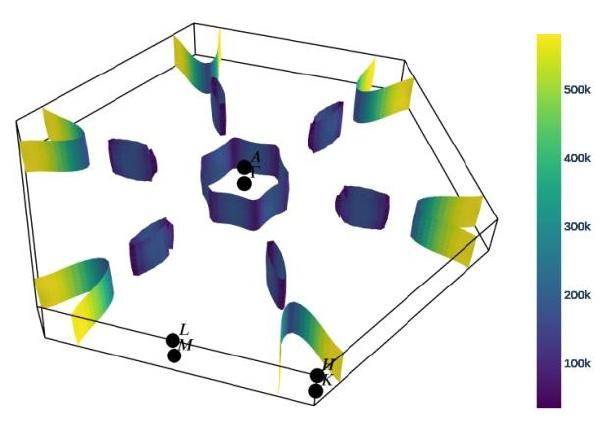}} \\
\subfigure[$ $]{\hspace{-0.8cm}\includegraphics[trim=0mm 4mm 0mm 3mm,clip,scale=0.63]{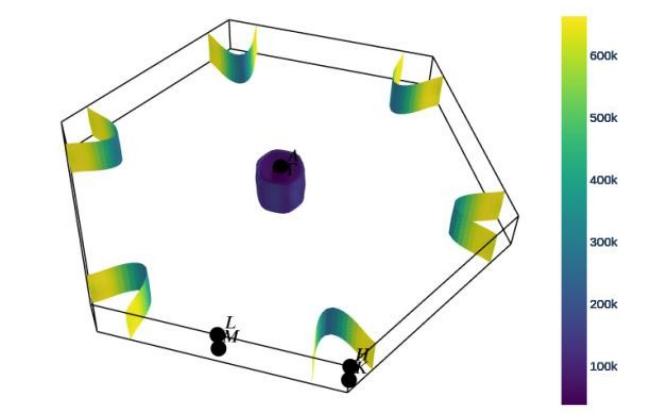}} \\
\subfigure[$ $]{\hspace{0.5cm}\includegraphics[trim=0mm 4mm 0mm 3mm,clip,scale=0.62]{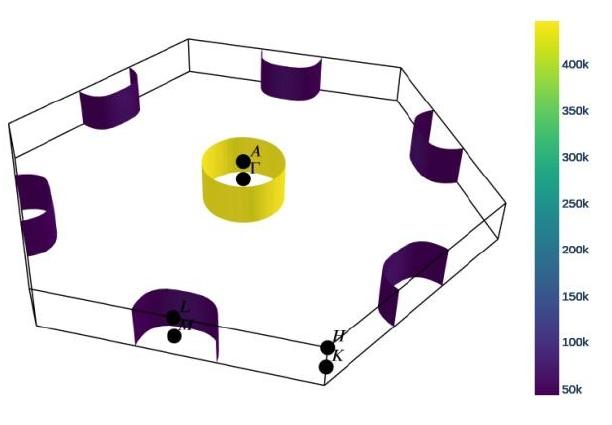}}
\subfigure[$ $]{\includegraphics[trim=0mm 4mm 0mm 0mm,clip,scale=0.62]{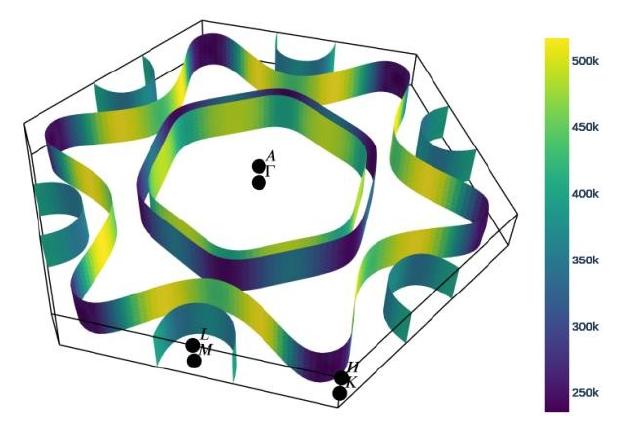}}
 \caption{Fermi surface of Janus MXenes (a) \ce{TiCrCO2}, (b) \ce{TiCrCF2}, (c) \ce{TiCrC(OH)2}, (d) \ce{MoCrCO2}, and (e) \ce{MoCrCF2}.}
 \label{FS}
\end{figure*}

 For a basic understanding of the carrier transport in these materials, we determine the Fermi surfaces and Fermi velocities of these conducting materials. The atom-resolved band structures of stable Janus MXenes are given in Figure S5-S6 and the corresponding Fermi surfaces are presented in Figure \ref{FS}. The different colors on the Fermi sheets correspond to the variation in Fermi velocity. In the spin-down channel of \ce{TiCrCF2}, there are two bands crossing the Fermi energy along the K - $\Gamma$ high symmetry direction. These bands originate from the admixture of Cr and Ti states, with Ti being dominant. The Fermi sheet corresponding to these bands has a punctured pipe like shape in Figure \ref{FS}(b) and low Fermi velocity. At K, there are semi-ellipsoids with pointed edges which are hexagonally symmetric with maximum Fermi velocity at the broad center which decreases towards the edge. Similar, semi-ellipsoids are seen in the case of \ce{TiCrC(OH)2}.

 In \ce{MoCrCO2}, there is a cylindrical Fermi sheet centered at $\Gamma$ corresponding to the hole pocket in spin down channel of band structure. It has relatively large Fermi velocity given by the yellow color on the velocity scale. There are no other bands crossing the Fermi level. In spin up channel, a band with major contribution from Cr leads to flat semi-circular Fermi sheets around M with low Fermi velocity. \ce{TiCrCO2} has similar Fermi surface feature at M. But, \ce{MoCrCF2} has semi-circular Fermi sheets corresponding to hole pocket at M. It is a metal with more bands crossing the Fermi energy leading to complex features enclosing large area in the Fermi surface.
\begin{table*}
\small
 \caption{Film conductivity of Janus MXenes}
 \begin{tabular*}{\textwidth}{@{\extracolsep{\fill}}|cccc|}
\hline
           & Fermi surface area & Fermi velocity (avg.) & Film conductivity \\ 
           &    \AA$^{-2}$            &   ($\times 10^{5} $  m/s)                 & ($\times 10^{5}$  S/m)  \\ \hline
\ce{TiCrCO2}    & 2.63 & 1.81 & 1.52  \\ 
\ce{TiCrCF2}    & 2.45 & 2.77  & 2.17  \\ 
\ce{TiCrC(OH)2} & 1.46 & 3.76 & 1.75  \\ 
\ce{MoCrCO2}    & 1.31 & 1.05 & 0.37  \\ 
\ce{MoCrCF2}    & 5.16 & 3.74 & 5.18  \\ 
\hline
\end{tabular*}
\label{film_conductivity}
\end{table*}
The Fermi surface area, velocity and electrical conductivity of Janus MXenes are calculated and presented in Table \ref{film_conductivity}. For the dynamically unstable \ce{MoCrC(OH)2}, these quantities are not calculated. For the semiconducting spin states, the electrical conductivity is taken to be zero. Among the five MXenes considered, \ce{MoCrCF2} has the highest film conductivity which can be attributed to large Fermi surface area and Fermi velocity. Though metallic \ce{TiCrC(OH)2} has a similar value for Fermi velocity, the Fermi surface area enclosed is comparatively smaller. Also, it can be seen that the fluorine terminated MXene has the highest value of conductivity in the first series (\ce{TiCrCT2}). It is to be noted that the value of conductivity changes with the interlayer separation as discussed in the following. When a MXene is formed from MAX phase by etching out "A" layer, there is expansion of d (interlayer separation), which is a characteristic of MAX to MXene transformation \cite{naguib2012two, naguib2013new}. This is more pronounced when LiF is used for etching, due to the intercalation of water and Li between the carbide layers \cite{ghidiu2014conductive, ghidiu2016ion, dillon2016highly}, which implies that d depends on the method of sample preparation (specially on the etchant used for synthesis). In experiments, it is common to tune the interlayer separation with annealing. For instance, in the case of Mo$_{2}$CT$_{z}$ films, Halim et al. reported that annealing reduces the interlayer spacing \cite{halim2018variable}. Therefore, the films are annealed to enhance the electrical conductivity as the interlayer separation and conductivity are inversely proportional to each other. The annealing temperature, also has an effect on conductivity.

\subsubsection{Magnetic Exchange interaction, MAE and Curie Temperature, $T_c$}

Having studied the electronic structure of MCrCT$_2$ Janus MXene, we proceed toward the detailed analysis of the  magnetic ground state. Two important parameters for evaluating the characteristics of magnetic materials are $T_c$ and MAE. In order to determine the potential of magnetism and the magnetic ground state for monolayer Janus MXenes, we calculate the total energies for ferromagnetic (FM), antiferromagnetic (AFM), and nonmagnetic (NM) states. Figure-S3 (a-f) demonstrates the creation of the ferromagnetic (FM) and two distinct AFM layouts. We compute the magnetic interaction parameter using a 3 $\times$ 2 $\times$ 1 supercell.
By comparing their respective energies, we find that the ferromagnetic configuration is the ground state (with the lowest energy) for all MXenes considered.
 To examine the exchange interaction, we use the Heisenberg model, whose Hamiltonian can be defined as follows.

\begin{equation}
H = - \sum_{i,j} J_1( M_i .M_j) - \sum_{k,l} J_2( M_k .M_l)   
\end{equation}

where $J_1$ and $J_2$ represents the intralayer first-nearest and second nearest exchange coupling parameter, respectively. The first nearest site pair is (i, j), (k,l) is the second nearest site pair and M is the net moment at the Cr sites. As previously mentioned, the contribution to the magnetic moment of Cr is taken into account as Ti/Mo exhibit no influence.  According to the model, shown in Figure-S3(a-f) exchange coupling parameter $J_1$ and $J_2$ can be represented as $\frac{2E_{tot}(AFM_2)- E_{tot}(AFM_1)- E_{tot}(FM)}{8M^2} $ and $\frac{E_{tot}(AFM_1)- E_{tot}(AFM_2)}{4M^2}$ respectively. It is possible to represent the total energy of these magnetic structures $E_{tot}(FM)$, $E_{tot}(AFM_1)$ and $E_{tot}(AFM_2)$ as:
\begin{align}
  E_{tot}(FM) &= -6 J_1M^2 - 6 J_2M^2 + E_0 \\
  E_{tot}(AFM_1) &= 2J_1M^2 + 2J_2M^2 + E_0 \\
  E_{tot}(AFM_2) &= 2J_1M^2 -2J_2M^2 + E_0
\end{align}

where $E_0$
 is the total energy excluding the magnetic coupling that is not sensitive to different magnetic
states. Effective exchange parameters are frequently used to describe the strength of the exchange interactions in a magnetic system, defined as $J_{eff}$ =  $\displaystyle \sum_{i\ne 0} J_i$, with i = 1,2.
$J_{eff}$ $<$ 0 and $J_{eff}$ $>$ 0 are two separate instances to think about while choosing the Hamiltonian. The ferromagnetic scenario favours parallel magnetic moments when $J_{eff}$ $>$ 0; this is because of the energy of their interaction. Instead, antiparallel orientations are preferred for $J_{eff}$ $<$ 0; this is the case for antiferromagnetism. For the current situation, the computed exchange interaction parameter, $J_i$, (Table-\ref{tab:mag1}) for all Janus structures show that the system favours the ferromagnetic structure. The estimated Curie temperature of the FM state is reported in Table-\ref{tab:mag1}, which is calculated using the mean field approximation as\cite{pajda2001ab} $T_c$ = $\frac{2M(M+1) }{3 k_B}$ $\displaystyle \sum_{i\ne 0} J_i$. The Curie temperature of TiCrCO$_2$ is the highest among all examined Janus MXenes, according to the estimated value.

\begin{figure}
  \centering
     \includegraphics[scale=0.4]{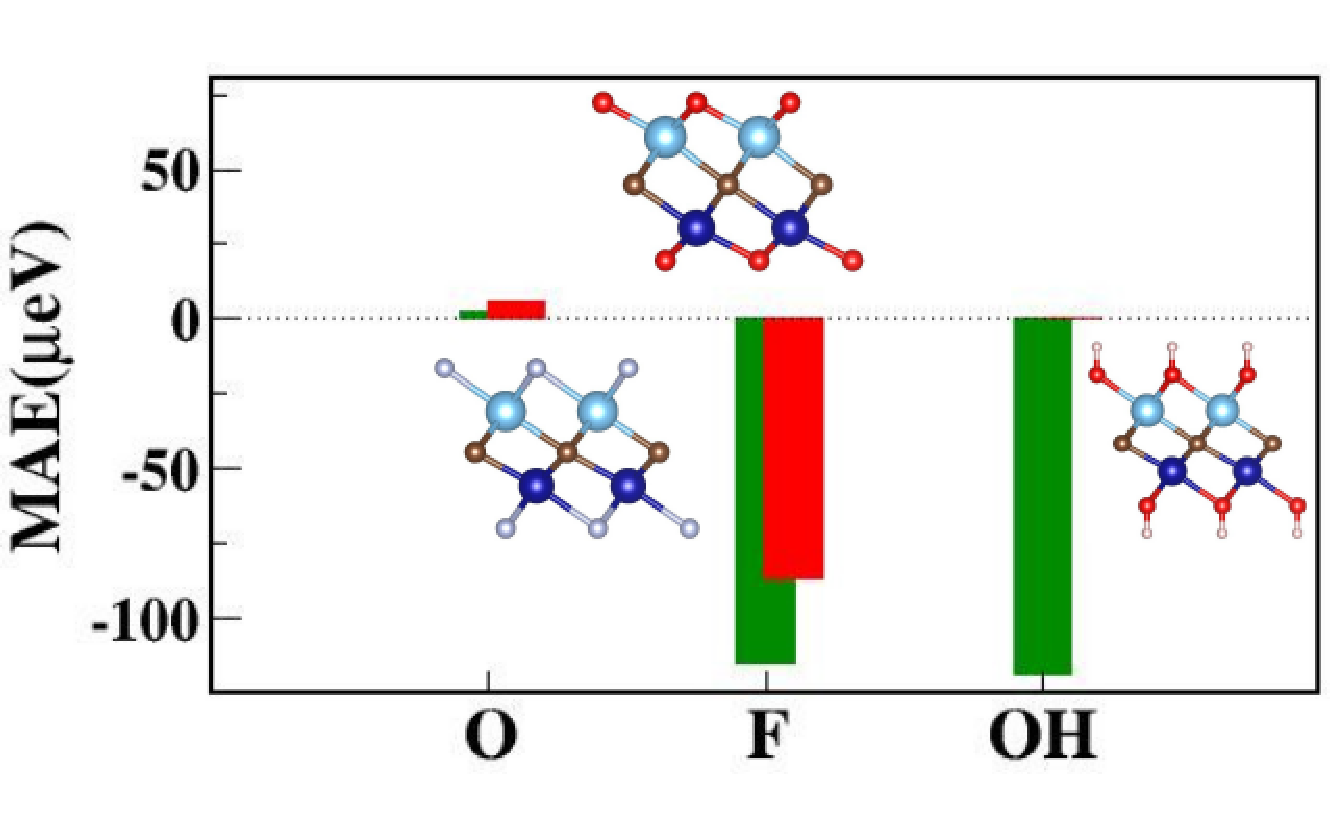}
 \caption{Bar plot of MAE of Janus MXene. Green bar represents TiCrCT$_2$ and red bar represent the MoCrCT$_2$ Janus MXene. }
 \label{Fig:MAE}
\end{figure}

Usually, the MAE is the sum of the contribution due to magneto-crystalline anisotropy energy (MCA) driven by SOC and the magnetic shape anisotropy energy (MSA) due to the magnetic dipole-dipole interactions~\cite{fan2023two,li2022electron} (MAE = MCA + MSA). As in two-dimensional systems, the latter (MSA) is much small in comparison to MCA, the SOC is the primary determinant of magnetic anisotropy, i.e. MAE $\approx$ MCA. We have computed MAE using the  total energy differences given by, 
\begin{align}
\begin{aligned}
MAE &= E_{[100]} - E_{[001]}    \\
\end{aligned}
\end{align}
Where $E_{[001]}$ and $E_{[100]}$ are the total energies of the structure with the magnetization in the [001] and [100] directions, respectively. In-plane magnetic anisotropy (IMA) and perpendicular magnetic anisotropy (PMA) are represented by the negative and positive MAE, respectively. From the Figure-\ref{Fig:MAE},
it is clear that O-terminated Janus MXenes favors the perpendicular magnetization direction (PMA), while -F and -OH terminated Janus MXenes exhibit IMA. Thus, M-O1, Cr-O2 has an exceptionally strong effect on switching the preferred magnetization axis of these materials from in-plane to out-of-plane.

\begin{table*}
\caption{The exchange coupling parameter per unit cell, $J_1$, $J_2$, magnetic anisotropic energy, MAE, magnetocrystalline
anisotropy, MCA and the Curie temperature, T$_c$ are represented here. }
 \begin{tabular*}{\textwidth}{@{\extracolsep{\fill}}|cccccc|}
\hline
           & $J_1$ & $J_2$  & MAE  & T$_c$   \\ 
            & (meV/unit cell)  & (meV/unit cell) &$\mu$eV  & (K)    \\ 
           \hline

TiCrCO$_2$     & 6.27  & 3.47 & 3  & 150.7 \\ 
TiCrCF$_2$    & 4.41    & 0.82 & -115  & 80.8 \\ 
TiCrC(OH)$_2$ & 4.05    & 0.896 & -119  &  76.5 \\ 
MoCrCO$_2$    &  3.47    & 0.137 & 6  & 55.8 \\ 
MoCrCF$_2$    & 5.37   & -0.27 & -87  & 78.9 \\ \hline
\end{tabular*}
\label{tab:mag1}
\end{table*}

\section{Micromagnetic Simulation of Janus MXene Monolayers}
To demonstrate that the Janus MXenes with different terminations show different magnetic switching behavior we performed micromagnetic simulation using UBERMAG framework~\cite{ubermag}, a comprehensive micromagnetic simulation tool, is employed for these simulations. Studying current-induced magnetization switching in these magnetic Janus monolayers offers significant advantages over traditional magnetic field-induced switching~\cite{bhattacharjee2012theoretical} as such an approach is particularly beneficial because it allows for more efficient manipulation of magnetic states, potentially leading to lower power consumption and higher operational speeds in spintronic devices. 
The total energy of the system is given by,\begin{equation}
E_{tot}=-A \mathbf{M} \cdot \nabla^2 \mathbf{M}-\frac{1}{2} \mu_0 M_{\mathrm{s}} \mathbf{M} \cdot H_d-K(\mathbf{M} \cdot \mathbf{u})^2
\end{equation}
The first term is the contribution from the exchange interaction, which is associated with the exchange stiffness \(A\), and the second term is the energy corresponding to the demagnetization field (${\bf H}_d$) The third term is due to the uniaxial anisotropy, which is characterized by the uniaxial anisotropy constant \(K\).
The current-induced switching was demonstrated using a spin transfer torque within the framework of Zhang-Li model~\cite{li2003magnetization}.  The magnetization dynamics are described by the modified Landau-Lifshitz-Gilbert equation which carries an STT term~\cite{li2003magnetization,bhattacharjee2023spin},
\begin{equation}
\begin{aligned}
\frac{d{\bf M}}{dt} = & -\frac{\gamma_0}{1+\alpha^2} \mathbf{M} \times \mathbf{H}_{\mathrm{eff}} - \frac{\gamma_0 \alpha}{1+\alpha^2} \mathbf{M} \times\left(\mathbf{M} \times \mathbf{H}_{\mathrm{eff}}\right) \\
& -(\mathbf{u} \cdot \boldsymbol{\nabla}) \mathbf{M} + \beta \mathbf{M} \times[(\mathbf{u} \cdot \boldsymbol{\nabla}) \mathbf{M}]
\end{aligned}
\label{Z-L}
\end{equation}

In the above equation, the first term in the r.h.s describes the precessional dynamics under the effective field ($\mathbf{H}_{\mathrm{eff}}$), while the second term is the damping term. The third term represents the spin-transfer torque.

The setup involves defining physical constants and material-specific parameters:

\begin{itemize}
    \item Standard physical constants include the vacuum permeability (\(\mu_0\)) and the electron charge (\(e\)).
    \item The Bohr magneton (\(\mu_B\)) and the gyromagnetic ratio (\(\gamma\)) are pivotal in magnetic calculations.
    \item Exchange energies for nearest (NN) and next-nearest neighbors (NNN), denoted as \(J_1\) and \(J_2\), are converted from meV to Joules to maintain consistency with SI units.
    \item The uniaxial anisotropy constant (\(Ku\)) is similarly converted from meV to Joules.
    \item The lattice constant (\(a\)) is assumed to be in Angstroms and is converted to meters.
\end{itemize}
The exchange stiffness constant \(A\) is calculated considering both NN and NNN interactions, with coordination numbers \(z_{NN} = 6\) and \(z_{NNN} = 6\). The employed relationship is \(A = \frac{12}{a}(J_1 + J_2)\), reflecting the hexagonal symmetry and the cumulative effect of NN and NNN interactions (a being the lattice constant).

We have used Gilbert damping constant \(\alpha=0.01\). The initialization of magnetic moments is contingent on the MXene surface termination: for -O terminated MXenes, the moments are aligned along the z-direction, while for -F and -OH terminated MXenes, they are oriented along the x-direction, consistent with the discussions in the previous section.
\begin{figure}
    \centering
    \includegraphics[width=0.75\linewidth]{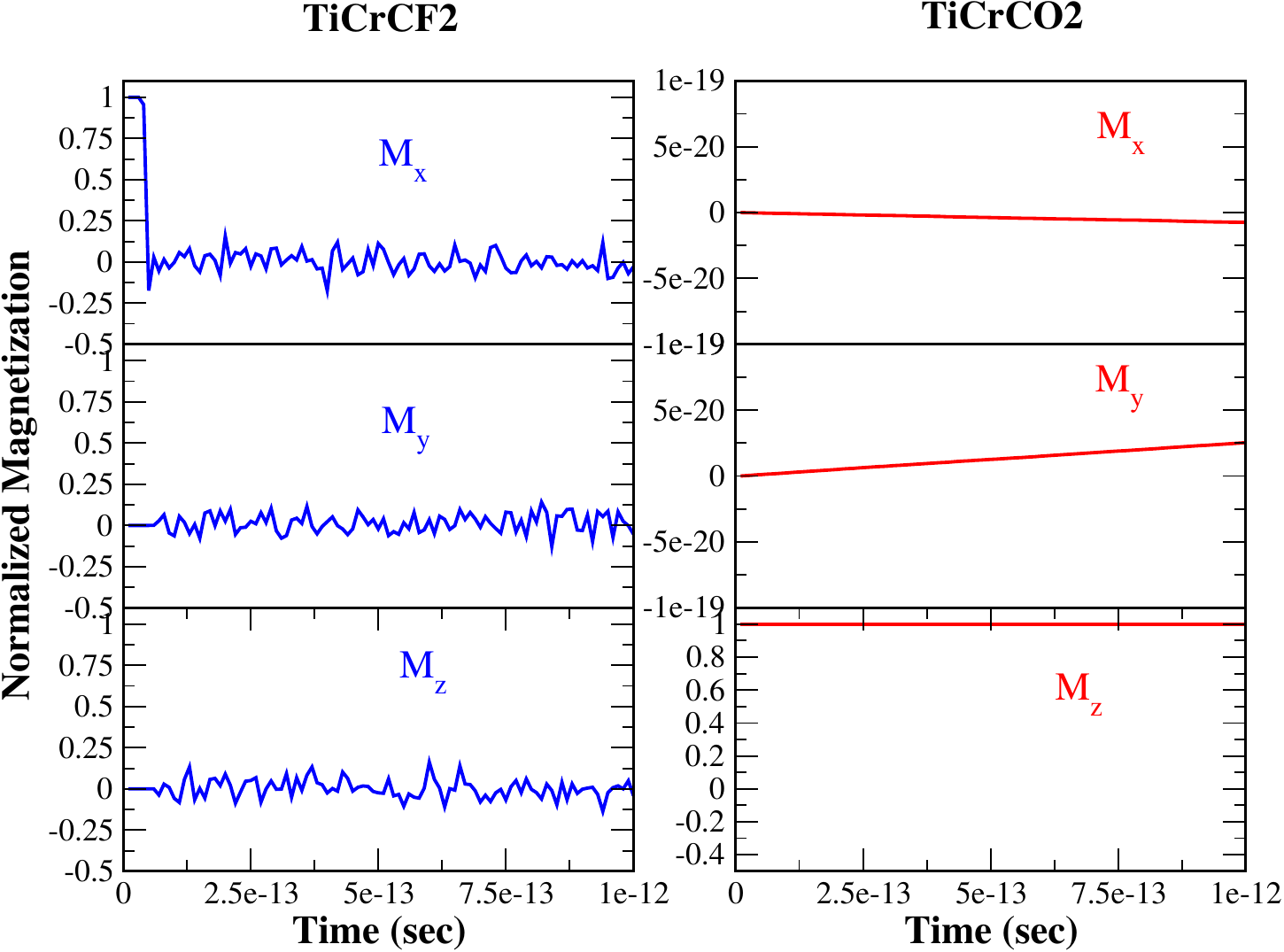}
    \caption{Illustrative example of micromagnetic simulation in TiCrCF$_2$ (left) and TiCrCO$_2$ (right) using Jhan-Li type of STT mechanism. Simulation is done using parameters $u=10^6$ m/s and non-adiabatic STT parameter $\beta$=0.5}
    \label{fig:STT}
\end{figure}
In Figure \ref{fig:STT}, we present a comparative analysis of the normalized magnetization components ($M_x$, $M_y$, $M_z$) as a function of time for two distinct materials: TiCrCF$_2$ and TiCrCO$_2$. This illustration depicts the dynamic behavior of the magnetization components under identical simulation conditions, with the velocity of injected carriers set at $10^6$ m/s for both materials.

For TiCrCF$_2$, where the easy axis is aligned along the [100] direction, the magnetization shows a rapid decline to zero, followed by a slight negative excursion within the first 0.25 picoseconds. This quick response indicates a swift reaction to the applied conditions.

In contrast, TiCrCO$_2$, with its easy axis along the [001] direction, demonstrates a markedly different behavior. Despite the application of a high current, the magnetization exhibits negligible variation throughout the simulation (1 picosecond). This observation suggests a higher threshold for achieving noticeable magnetization switching in TiCrCO$_2$, potentially due to its intrinsic material properties or the orientation of its easy axis.
The distinct magnetization dynamics of TiCrCF$_2$ and TiCrCO$_2$ can be elucidated through the Zhang-Li model, as described by the given equation (Eq.\ref{Z-L}). In this model, several terms collectively influence the magnetization behavior, including the precession around the effective magnetic field $\mathbf{H}_{\mathrm{eff}}$, the damping effect proportional to $\alpha$, and the spin-transfer torque terms influenced by carrier velocity $\mathbf{u}$. The stronger exchange energy in TiCrCO$_2$ suggests a more robust internal magnetic interaction, which could lead to greater resistance against changes in magnetization under identical external conditions, as compared to TiCrCF$_2$. The higher exchange energy in TiCrCO$_2$ may explain the minimal variation in its magnetization, even under a higher current, indicating a need for a higher threshold to achieve significant magnetization switching. Conversely, the relatively weaker exchange energy in TiCrCF$_2$ allows for a more rapid and noticeable response to the injected carrier velocity, aligning with the observed quick decline in magnetization. 
These findings highlight the significant impact of material composition, magnetic exchange interaction, and anisotropy on the magnetization dynamics in spintronic applications. The distinct responses of TiCrCF$_2$ and TiCrCO$_2$ under identical simulation parameters underscore the necessity for tailored approaches when designing spintronic devices.

\section{Conclusions}
In conclusion, the structural, electronic and magnetic properties of 2D Janus MXenes MCrCT$_2$ (M = Ti, Mo; T = O, F, OH) are studied. Density functional theory is used to examine a stable monolayer Janus MXene MCrCT$_2$ with inherent ferromagnetism. We have demonstrated that only five magnetic Janus structures—TiCrCO$_2$, TiCrCF$_2$, TiCrC(OH)$_2$, MoCrCO$_2$, and MoCrCF$_2$—are dynamically and thermodynamically stable, while the magnetic Janus type MoCrC(OH)$_2$ is unstable. Only TiCrCO$_2$ and TirCF$_2$ have a half metallic character with 100$\%$ spin polarisation. By analysing the electrostatic potential, the potential difference between the two surface layers of Janus MXene is verified. All predicted stable Janus MXenes are ferromagnetic since the exchange coupling values have been determined to be positive. According to the MAE calculation, Janus structures with -O terminations have perpendicular magneto anisotropy, but Janus MXenes with -F and -OH terminations have in plane magneto anisotropy.
 From the evaluation of Fermi velocity and Fermi surface area, we find that these MXenes have relatively large electrical conductivity, which is beneficial for its potential application in microelectronics.

\section*{Conflicts of interest}
The authors declare that they have no conflict of interest.

\section*{Acknowledgements}
This work was supported by NRF grant funded by MSIP, Korea 
(No. 2009-0082471 and No.
2014R1A2A2A04003865), the Convergence Agenda Program (CAP) of the Korea Research Council of 
Fundamental Science and Technology (KRCF)and GKP (Global Knowledge Platform) project of the 
Ministry of Science, ICT and Future Planning. The authors would also like to acknowledge support from the Korea Institute of Science and Technology Information (KISTI) Supercomputer Center through their RnD innovation support program (Grant No. KSC-2022-CRE-0510).



\renewcommand\refname{References}



\balance


\bibliography{rsc} 
\bibliographystyle{rsc} 

\end{document}


\centering
\noindent\large{\textbf{Electronic Supplementary Information}} \\
\vspace{0.2cm}
\noindent\LARGE{\textbf{Tuning the  Electronic and Magnetic Properties of Double Transition Metal MCrCT$_2$ (M = Ti, Mo) Janus MXenes for Enhanced Spintronics and Nanoelectronics}} \\
\vspace{0.3cm} 

\noindent\large{Swetarekha Ram,\textit{$^{a}$} Namitha Anna Koshi,\textit{$^{a}$} 
Seung-Cheol Lee$^{\ast,}$\textit{$^{a,b}$} and Satadeep Bhattacharjee$^{\ast,}$\textit{$^{a}$}} \\
\vspace{0.3cm} 
 
\noindent{\textit{$^{a}$Indo-Korea Science and Technology Center (IKST), Jakkur, Bengaluru 560065, India}} \\
\noindent{\textit{$^{b}$Electronic Materials Research Center, KIST, Seoul 136-791, South Korea}} \\
\vspace{0.3cm} 
\noindent{$^{\ast}$E-mail: leesc@kist.re.kr, s.bhattacharjee@ikst.res.in} 
\vspace{1.5cm}

\FloatBarrier

\paragraph{Phonon dispersion:}
We presented the \ch{MoCrCOH2} phonon dispersion, which has an imaginary frequency and is not included in any investigation in the present case.
\begin{figure}
  \centering
     {\includegraphics[scale=0.5]{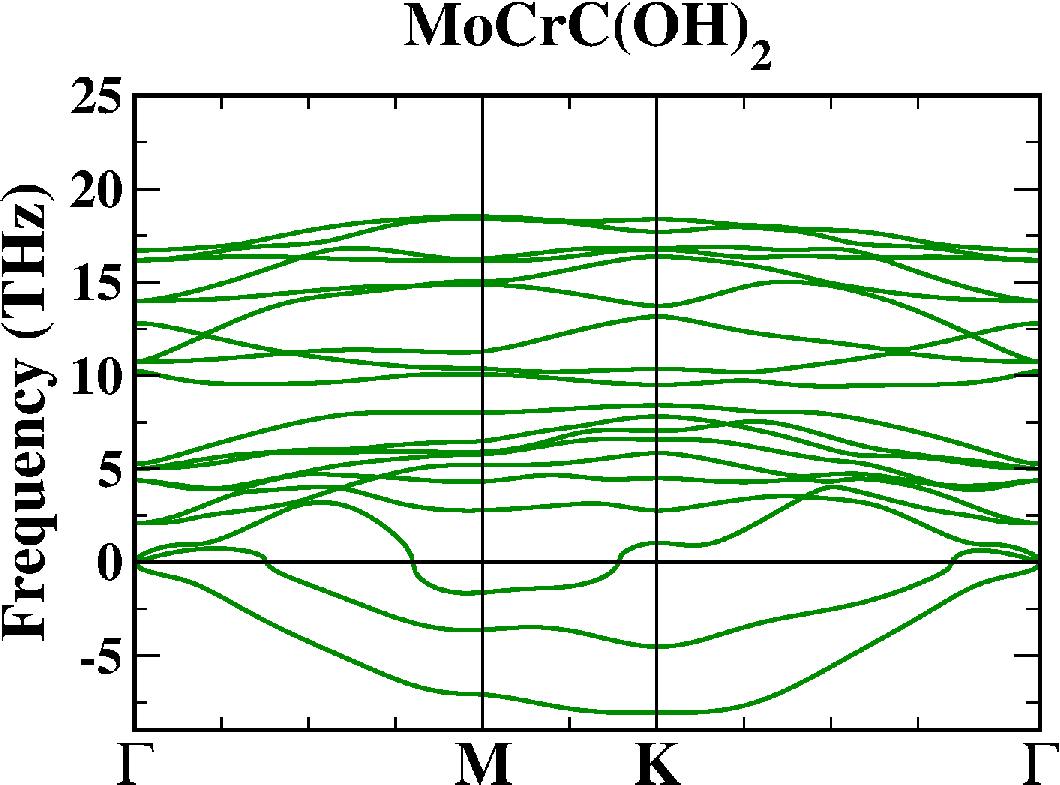}}
 \caption{Phonon dispersion of  MoCrC(OH)$_2$. The negative frequency indicates the dynamical instability of MoCrC(OH)$_2$. }
 \label{Fig:phonon}
\end{figure}

\begin{figure}
  \centering
\subfigure[]{\includegraphics[scale=0.35]{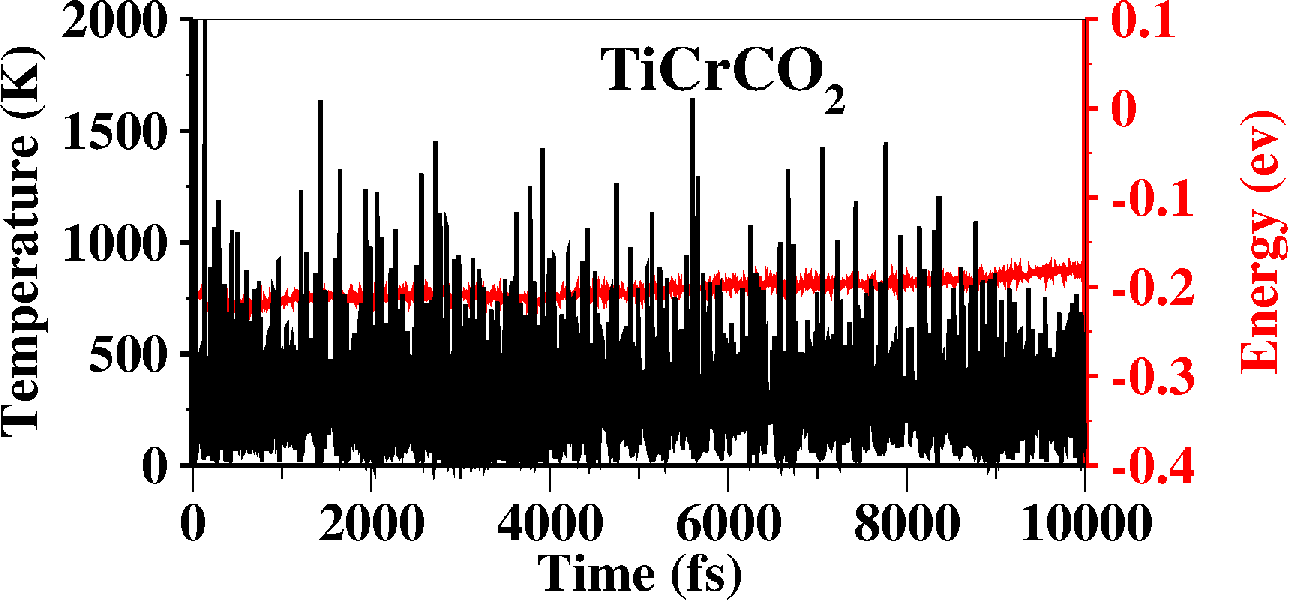}} 
     \subfigure[]{\includegraphics[scale=0.35]{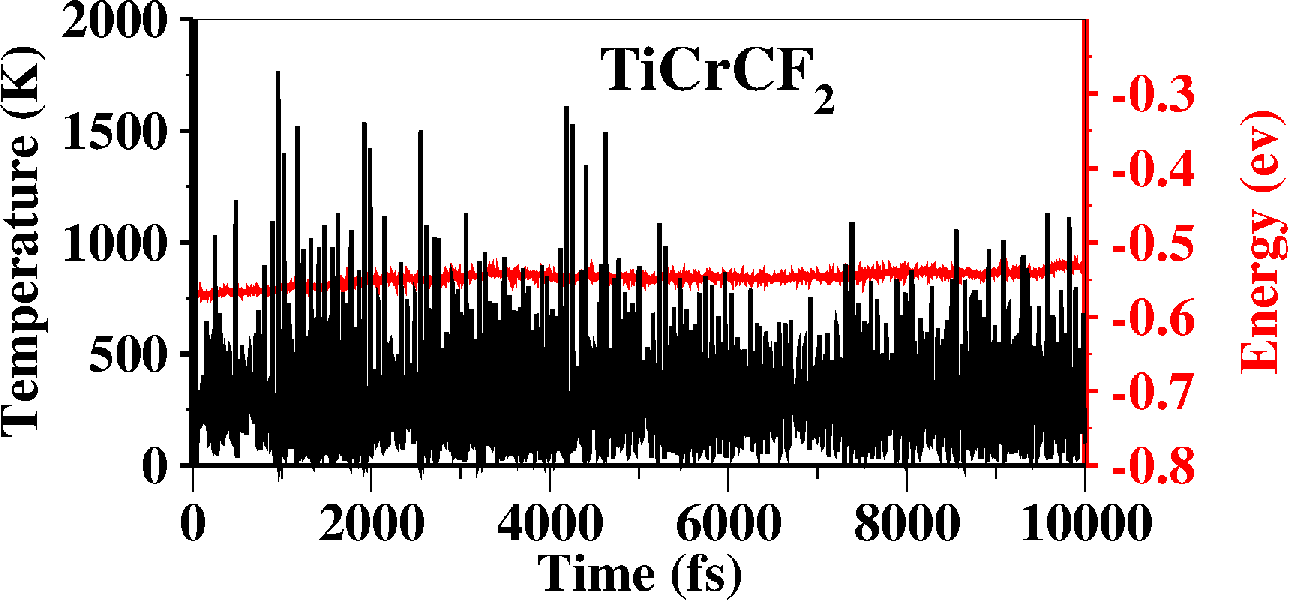}}  \\
     \subfigure[]{\includegraphics[scale=0.35]{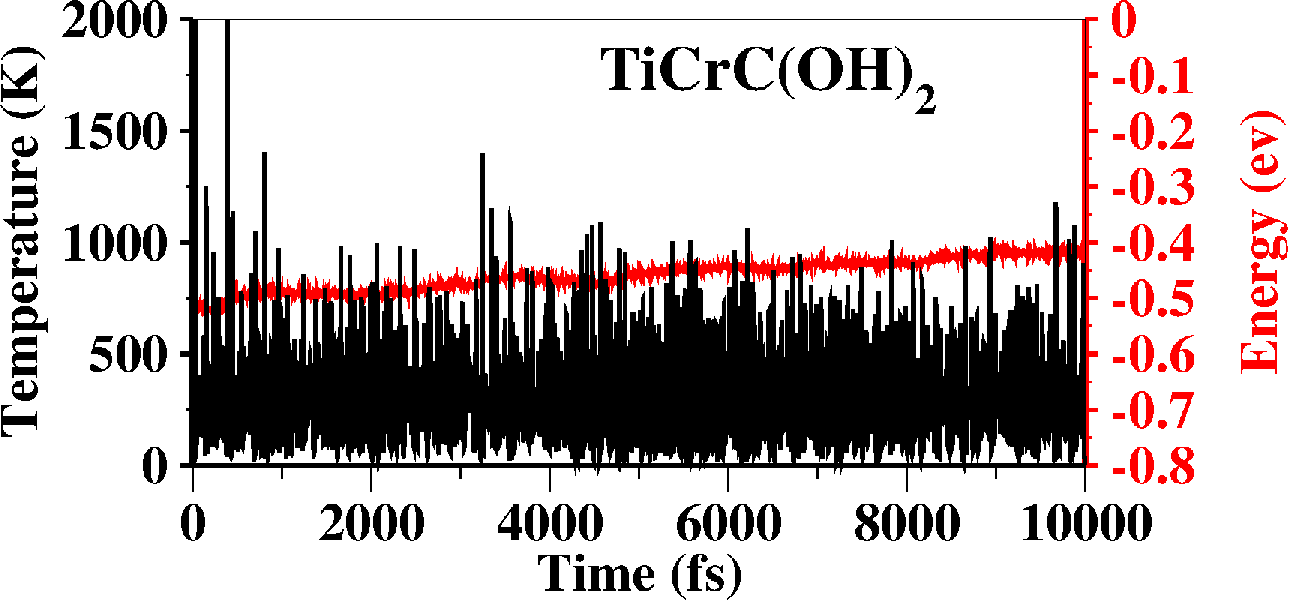}} \\ 
     \subfigure[]{\includegraphics[scale=0.35]{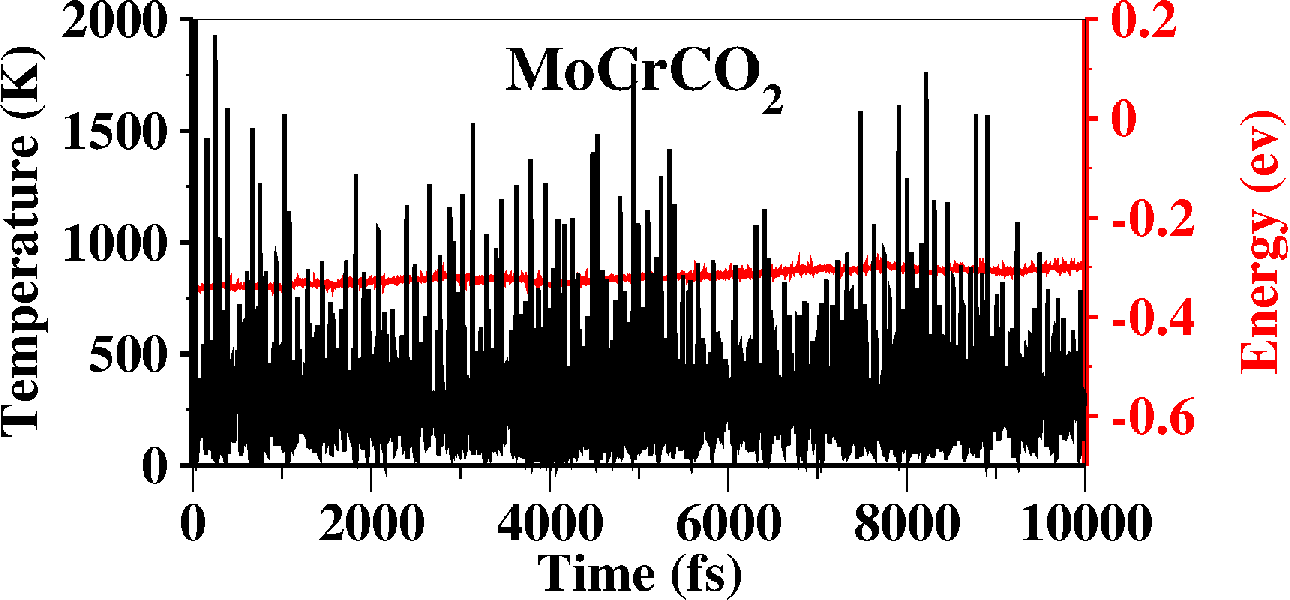}}  
     \subfigure[]{\includegraphics[scale=0.35]{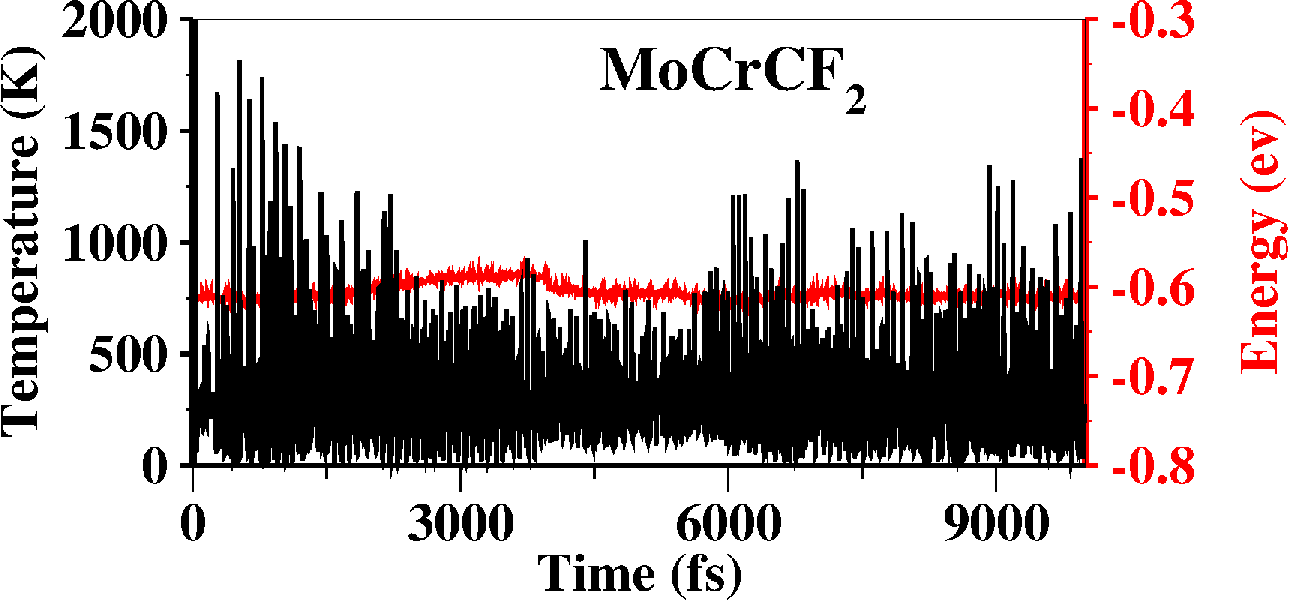}}
     \caption{Change of free energy (red line) and temperature (black line) over a time of 10000 fs in AIMD simulations at 300K. }
 \label{Fig:AIMD}
\end{figure}

\begin{figure}
  \centering
   \subfigure[]{\includegraphics[scale=0.33]{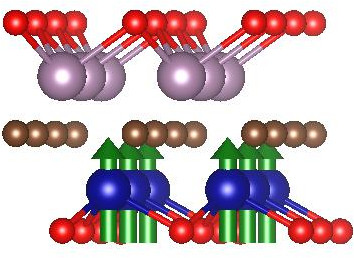}}
    \subfigure[]{\includegraphics[scale=0.18]{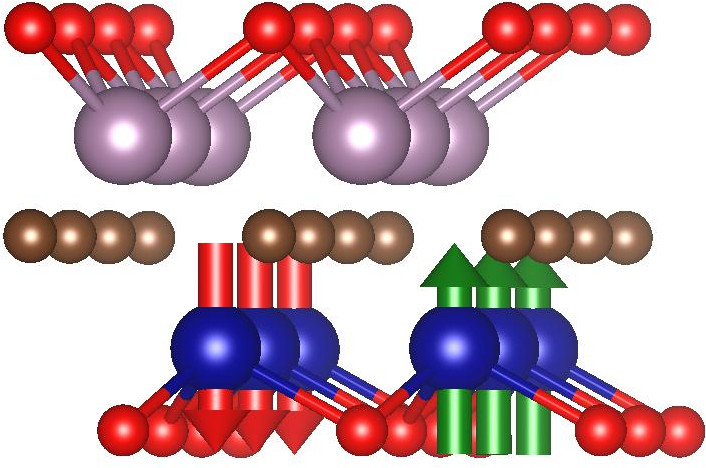}}
     \subfigure[]{\includegraphics[scale=0.33]{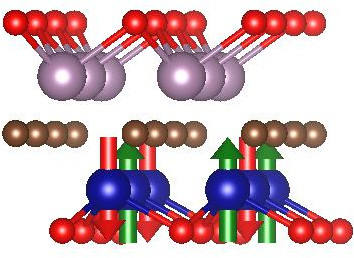}}\\
     \subfigure[]{\includegraphics[scale=0.25]{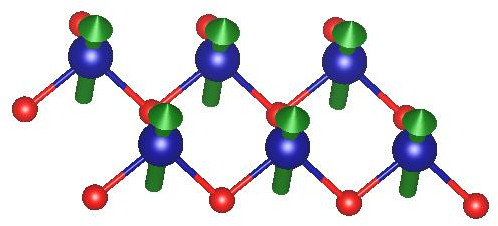}}
     \subfigure[]{\includegraphics[scale=0.35]{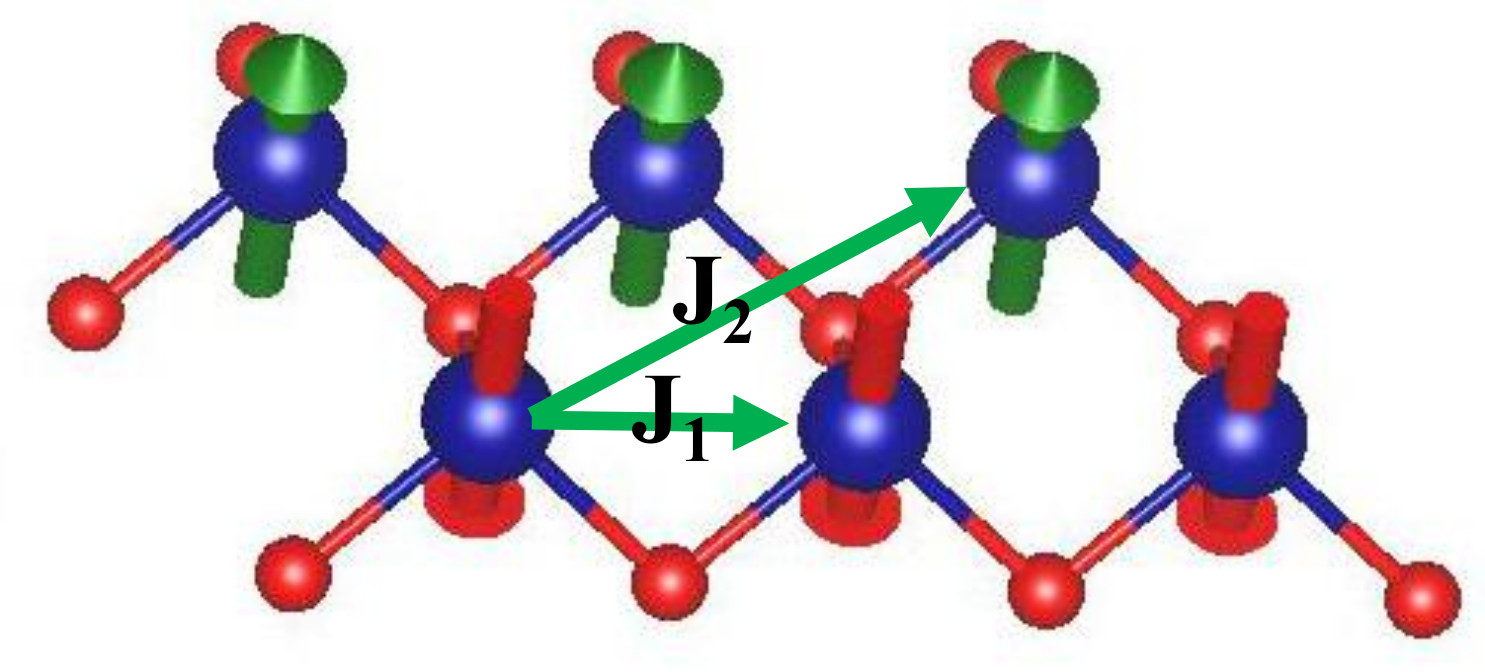}}
     \subfigure[]{\includegraphics[scale=0.25]{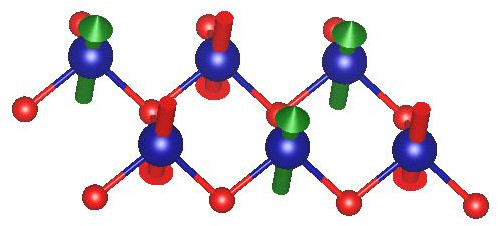}}
     
      \caption{Different orientations of Janus MXene configurations: (a,d) FM, (b,e) AFM1, (c,f) AFM2. Red and green arrows, respectively, indicate the up and down spins. The nearest and next nearest neighbours interactions are denoted by $J_1$ and $J_2$.}
 \label{Fig:magnotation}
\end{figure}


%
\begin{figure}
\subfigure[$ $]{\hspace{-1.5cm}\includegraphics[trim=0mm 8mm 0mm 0mm,clip,scale=0.65]{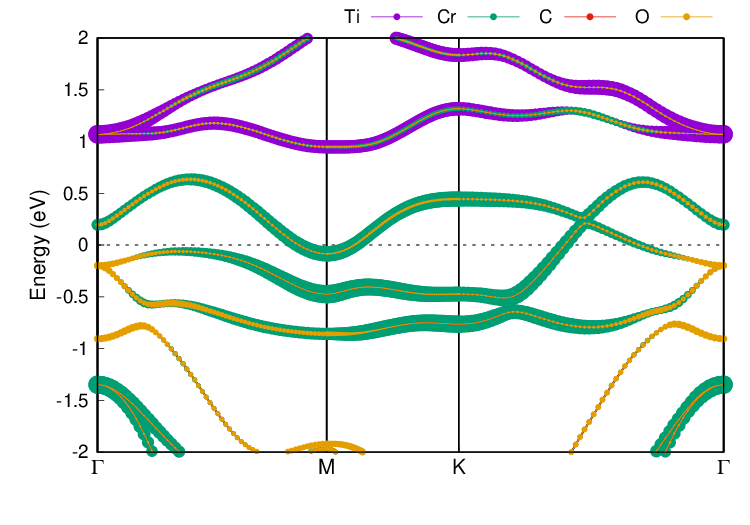}} 
\subfigure[$ $]{\hspace{0.5cm}\includegraphics[trim=0mm 8mm 0mm 0mm,clip,scale=0.65]{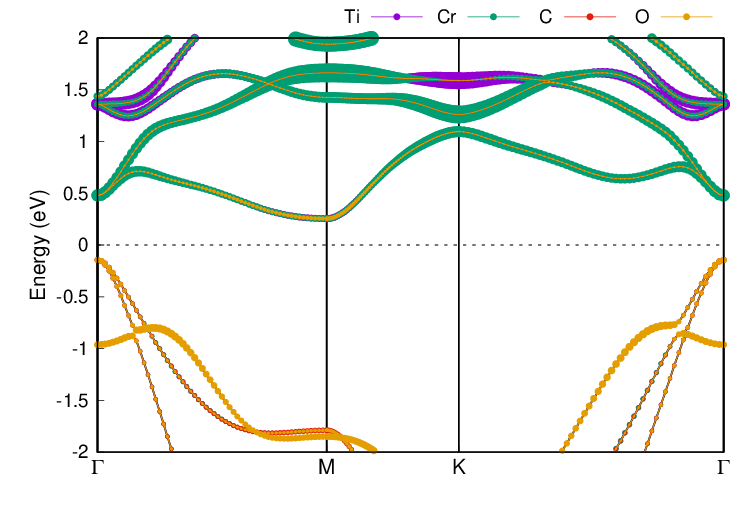}} 
\subfigure[$ $]{\hspace{-1.5cm}\includegraphics[trim=0mm 8mm 0mm 0mm,clip,scale=0.65]{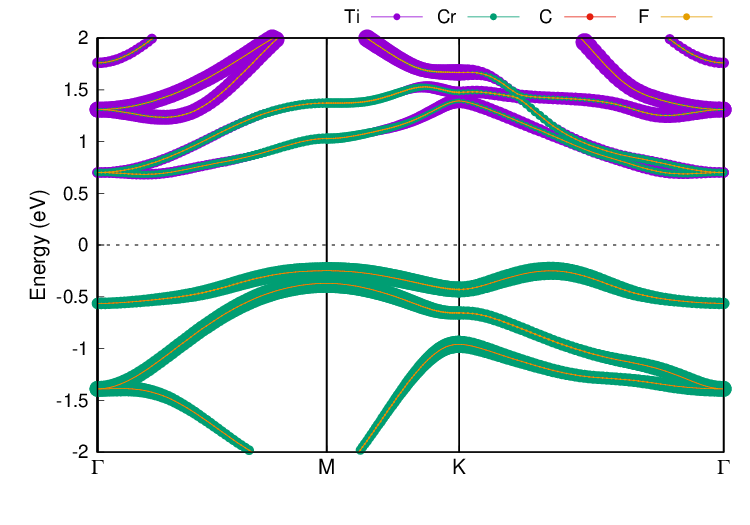}} 
\subfigure[$ $]{\hspace{0.5cm}\includegraphics[trim=0mm 8mm 0mm 0mm,clip,scale=0.65]{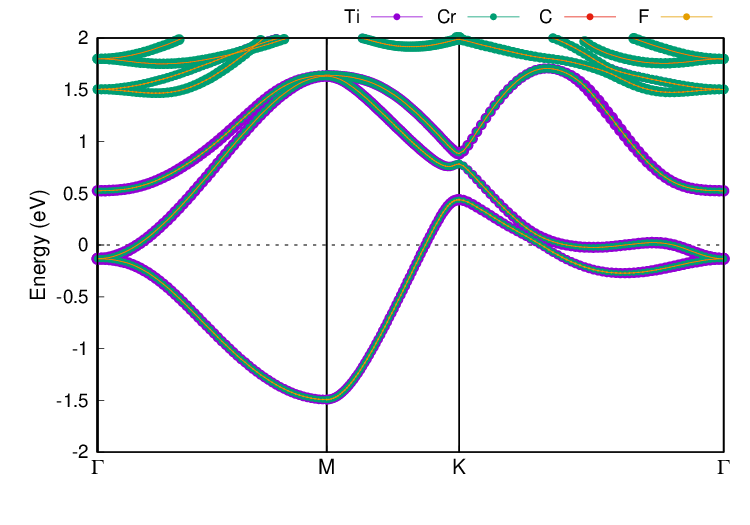}} 
\subfigure[$ $]{\hspace{-1.5cm}\includegraphics[trim=0mm 8mm 0mm 0mm,clip,scale=0.65]{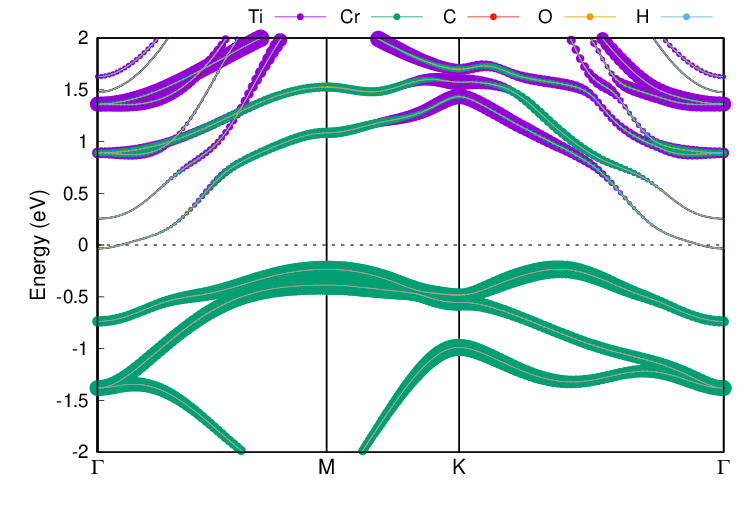}} 
\subfigure[$ $]{\hspace{0.5cm}\includegraphics[trim=0mm 8mm 0mm 0mm,clip,scale=0.65]{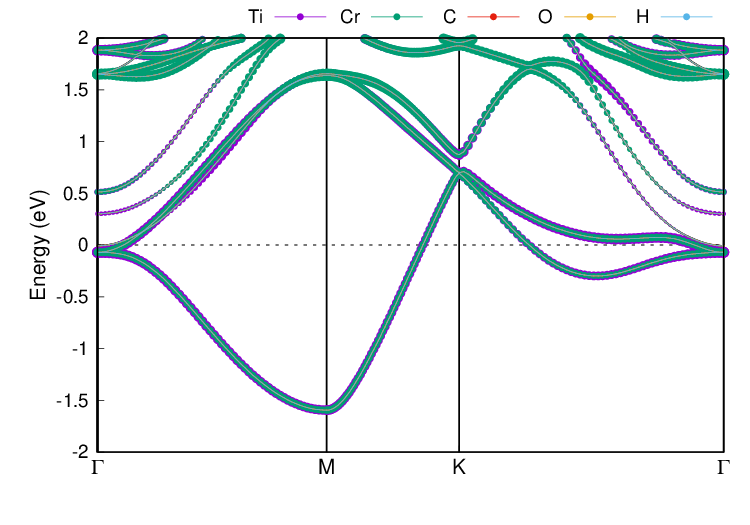}} 
 \caption{Atom resolved band structures of (a,b) \ch{TiCrCO2}, (c,d) \ch{TiCrCF2}, and (e,f) \ch{TiCrC(OH)2}. Spin up state is on the left for each MXene and spin down state is on the right.}
 \label{bnd_TiCr}
\end{figure}
\begin{figure}
\subfigure[$ $]{\hspace{-1.5cm}\includegraphics[trim=0mm 8mm 0mm 0mm,clip,scale=0.65]{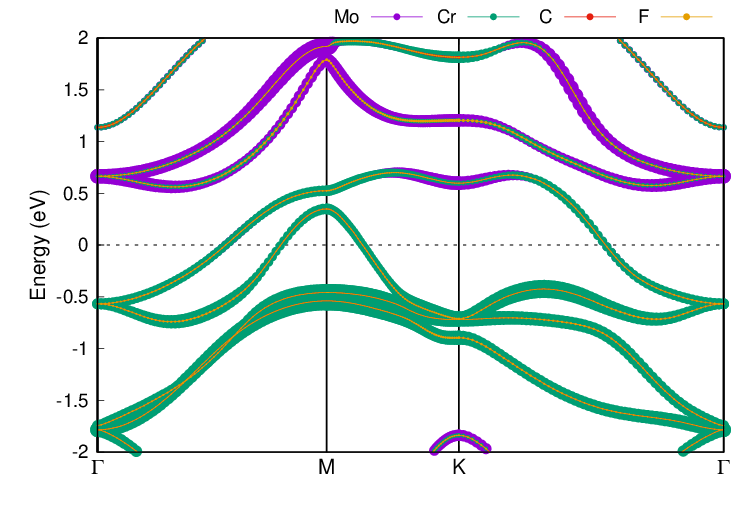}} 
\subfigure[$ $]{\hspace{0.5cm}\includegraphics[trim=0mm 8mm 0mm 0mm,clip,scale=0.65]{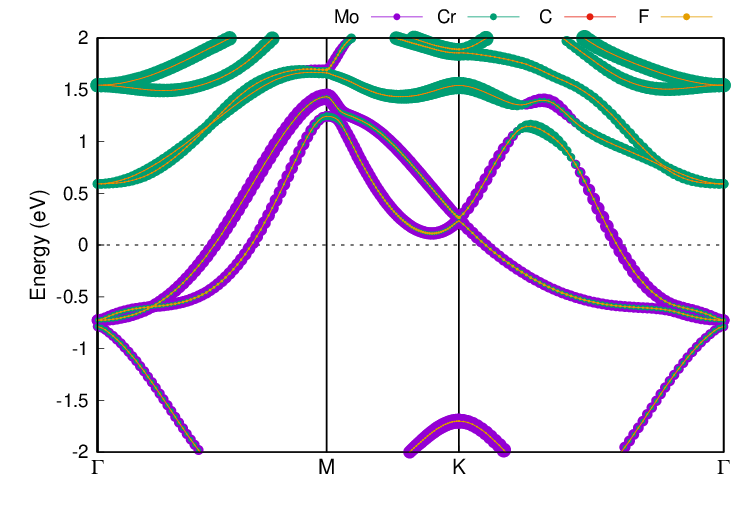}} 
\subfigure[$ $]{\hspace{-1.5cm}\includegraphics[trim=0mm 8mm 0mm 0mm,clip,scale=0.65]{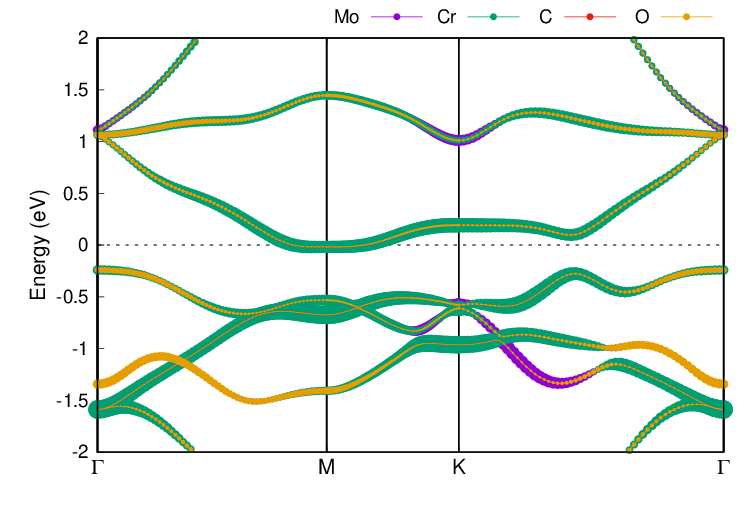}} 
\subfigure[$ $]{\hspace{0.5cm}\includegraphics[trim=0mm 8mm 0mm 0mm,clip,scale=0.65]{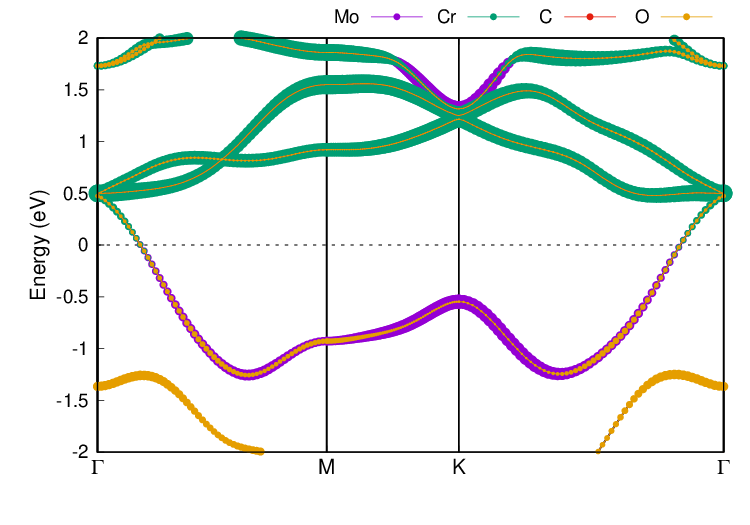}} 
 \caption{Atom resolved band structures of (a,b) \ch{MoCrCF2}, and (c,d) \ch{MoCrCO2}. Spin up state is on the left for each MXene and spin down state is on the right.}
 \label{bnd_MoCr}
\end{figure}

\FloatBarrier